\documentclass[reprint,secnumarabic,nobibnotes,aps,prx,superscriptaddress]{revtex4-2}
\usepackage[english]{babel}
\usepackage{color, xcolor, colortbl}
\usepackage{graphicx}
\usepackage{amsthm,amsmath,amssymb,amsfonts}
\usepackage{algorithm}
\usepackage{algorithmic}
\usepackage{dcolumn}
\usepackage{epstopdf}
\usepackage{bm}
\usepackage{subfigure}
\usepackage{appendix}
\usepackage{multirow}
\usepackage{braket}
\usepackage{hyperref}
\usepackage[capitalize]{cleveref}
\usepackage{xpatch}
\usepackage{tikz}
\usepackage{adjustbox}
\usepackage{xspace}

\providecommand{\definitionname}{Definition}
\providecommand{\assumptionname}{Assumption}
\providecommand{\corollaryname}{Corollary}
\providecommand{\lemmaname}{Lemma}
\providecommand{\propositionname}{Proposition}
\providecommand{\remarkname}{Remark}
\providecommand{\theoremname}{Theorem}

\usetikzlibrary{fit}
\tikzset{%
  highlight/.style={rectangle,rounded corners,fill=blue!15,draw,fill opacity=0.3,thick,inner sep=0pt}
}

\newtheorem{remark}{Remark}

\begin{document}

\title{Stochastic Schr\"{o}dinger equation approach to real-time dynamics of Anderson-Holstein impurities: an open quantum system perspective}

\author{Zhen Huang}
\affiliation{Department of Mathematics, University of California,
Berkeley, CA 94720, USA}

\author{Limin Xu}
\affiliation{Department of Mathematical Sciences, Tsinghua University, Beijing 100084, China}

\author{Zhennan Zhou}
\email{zhennan@bicmr.pku.edu.cn}
\affiliation{Beijing International Center for Mathematical Research, Peking University, Beijing,
100871, China}

\begin{abstract}
We develop a stochastic Schr\"{o}dinger equation (SSE) framework to simulate real-time dynamics of Anderson-Holstein (AH) impurities coupled to a continuous fermionic bath. The bath degrees of freedom are incorporated through fluctuating terms determined by exact system-bath correlations, which is derived in an {\it{ab initio}} manner. We show that such an SSE treatment provides a middle ground between numerically expansive microscopic simulations and macroscopic master equations. Computationally,
the SSE model enables efficient numerical methods for propagating stochastic trajectories. We demonstrate that this approach not only naturally provides microscopically-detailed information unavailable from reduced models, but also captures effects beyond master equations, thus serves as a promising tool to study open quantum dynamics emerging in physics and chemistry.
\end{abstract}
       \maketitle 

\section{Introduction}\label{sec:intro}
Real-time dynamics of quantum impurity systems with fermionic baths have been a central topic in condensed matter physics for the past few decades. It is used to model a wide range of systems, such as magnetic impurities in metals \cite{anderson1961localized}, quantum dot systems \cite{hanson2007spins}, atom adsorption onto surface \cite{brako1981slowly}. Among various types of quantum impurity models, the Anderson-Holstein (AH) model \cite{newns1969self,holstein1959studies} is of critical importance, since it is directly related to chemisorption \cite{newns1969self}, electrochemistry \cite{persson1993applications}, heterogeneous catalysis \cite{luo2016electron} and molecular junctions \cite{nitzan2003electron}.

The Anderson-Holstein model comprises of a molecular system as the impurity part and 
the continuous heat bath equilibrated at a certain temperature as the environment. For example, when modeling molecule-metal interfaces, the bath part is made up of metal orbitals of a continuum of spectra \cite{newns1969self,holstein1959studies}. The most straightforward simulation strategy would be a direct discretization of the bath orbitals (for example, see
\cite{shenvi2009nonadiabatic,huang2023efficient}), and then propagate the many-body Schr\"{o}dinger equation for the entire extended system. 
However, such calculations are often so expensive that either one is only capable of studying a model with a very small number of bath orbitals \cite{katz2005role},
or one makes crude single-particle approximations \cite{shenvi2006vibrational}. These treatments are far from exact and are only effective in limited scenarios.

The impurities could be viewed as a fermionic open quantum system due to the influence of the infinite bath.
To incorporate the open-system effects with a manageable computational cost, various master equations are developed based on different physical approximations.  This is generally seen in the study of molecular-metal interfaces. Treating the nuclei of the molecular system as classical particles, classical master equations (CME) \cite{langreth1991derivation,dou2016broadened} are proposed along with surface hopping technique \cite{ouyang2015surface,miao2019comparison} to efficiently sample the ensemble at equilibrium. However, due to the breakdown of the Born-Oppenheimer approximation, it is necessary to capture the nuclear quantum effects and nonadiabatic dynamics simultaneously. This is often achieved by quantum master equations (QME), which describe the dynamics for the reduced density matrix by tracing out the electronic degree of freedom.
Lindblad master equations \cite{gao1997dissipative,wang2020combining} are used when one makes the Markovian approximation. To capture the memory effects,
Nakajima-Mori-Zwanzig equations \cite{nakajima1958quantum,fay2021chirality}, Redfield equations \cite{redfield1965theory,leathers2006density} and other non-Markovian equations \cite{elste2008current,elste2010effect,elste2011transport} are proposed as effective models. QME is widely used to model realistic systems \cite{elste2011transport,elste2011transport2,elste2010effect,haertle2013decoherence}. The hierarchy quantum master equation (HQME) \cite{tanimura1989time,erpenbeck2019hierarchical,haertle2013decoherence}, also known as the hierarchy equation of motion (HEOM), is a nonperturbative alternative but is often computationally too expensive.

However, there is another widely-used approach, namely Stochastic Schr\"{o}dinger equations (SSE),  for studying open quantum systems. By incorporating stochastic fluctuations arising from interactions with the external environment, SSE has been successfully used in modeling quantum decoherence\cite{goetsch1996effect,shiokawa1995decoherence}, quantum measurement \cite{wiseman1996quantum,gambetta2002non,gambetta2003interpretation}, quantum jumps \cite{breuer1995stochastic,breuer2009stochastic} and so on. It is found to be consistent with QME calculations in many cases \cite{zhao2012fermionic,nathan2020universal,breuer2002theory,strunz2004convolutionless}, and in the meantime has the following additional advantages: on the one hand, SSE provides an ensemble of time-dependent quantum trajectories, and therefore is very convenient for studying the statistical and nonequilibrium properties of the open system; on the other hand,  the numerical methods for simulating stochastic differential equations have already been studied extensively \cite{jentzen2011taylor,zhang2017numerical}, and as a result the above trajectories could be obtained effectively with a Monte-Carlo sampling scheme. However, the success of SSE models highly relies on the accurate modeling of the environment, i.e. obtaining the exact time-correlation function of the noise term in SSE.

Even though SSE is a very powerful tool in the modeling of open quantum systems, it has not been applied to study the real-time dynamics of quantum impurities with fermionic bath, to the best of our knowledge. On the one hand, SSE approaches are mostly developed for bosonic heat bath in different regimes. On the other hand, although there are attempts of using SSE to study fermionic bath effects \cite{zhao2012fermionic,shi2013non}, there lacks an ab initio treatment of the stochastic and non-Markovian effects, therefore not applicable to general chemical systems such as molecule-metal interfaces.

The purpose of this article is two-fold. On the one hand, we aim to fill in a gap in the study of Anderson-Holstein models as an open quantum system:  the time-correlation function of the noise term in SSE is obtained directly from the Anderson-Holstein Hamiltonian. On the other hand, we emphasize  the following hierarchy of modeling: SSE is at an intermediate level of approximation, right between the atomic-level AH model and the  QME model:
 $$
 \text{Anderson-Holstein}\rightarrow {\mathrm{SSE}}\rightarrow \mathrm{QME}.$$
For a detailed description of this modeling hierarchy, please see \cref{fig:nMSSE}.
Although every SSE is associated with a corresponding QME, SSEs are able to encapsulate many intricate microscopic details, while QMEs merely describe the statistical averages of the SSE trajectories. This is also seen in our numerical experiments \cref{sec:numericalresults}. 
In other words, QME should be viewed as a further approximation on top of SSE.

The rest of this article is organized as follows. In \cref{sec:ah2sse}, we derive the stochastic Schr\"{o}dinger equations from the Anderson-Holstein model. After introducing the problem setup in \cref{subsec:setup}, we review the Bogoliubov transformation in \cref{subsec:Bogoliubov}, and then derive the time correlation function in \cref{subsec:timecorrelation}. We finally arrive at the non-Markovian SSE model in \cref{subsec:nMSSE} and its wide band limit in \cref{subsec:widebandlimit}. In \cref{subsec:derivation}, we discuss how to obtain quantum master equations from SSE by taking expectation values of density matrices. In \cref{subsec:manyqme}, we show how different approximations would lead to different versions of QME, and demonstrate the hierarchy of modeling in detail. In \cref{sec:numericalmethods}, we discuss the numerical methods for trajectory-based SSE simulations. We introduce how to generate time-correlated noise in \cref{subsec:noise}, and how to do time evolution in \cref{subsec:timeevol}. We show numerical experiments in \cref{sec:numericalresults}. We demonstrate that SSE offers directly an ensemble of particle trajectories with microscopic details, and compare its results to QME.

\section{From AH model to SSE}
\label{sec:ah2sse}
\subsection{Setup}
\label{subsec:setup}
In the Anderson-Holstein model \cite{holstein1959studies},
 the total Hamiltonian consists of three parts: the system Hamiltonian $\hat H_S$, the bath Hamiltonian $\hat H_B$, and the system-bath interaction $\lambda \hat{H}_{S-B}$:
\begin{equation}
\hat{H}=\hat{H}_S+\hat{H}_B+\lambda \hat{H}_{S-B}.
\end{equation}
Here $\hat{H}_S$ is the system Hamiltonian, $\hat{H}_B$ is the bath Hamiltonian, and $\lambda\hat{H}_{S-B}$ is the Hamiltonian of the interaction between the system and bath where $\lambda$ is the interaction strength. 
The system Hamiltonian $\hat{H}_S$ consists of the kinetic and potential energy of the nuclei:
\begin{equation}
    \hat{H}_S  =\frac{\hat{p}^2}{2 m_{\mathrm{n}}}+U_0(\hat{x})+h(\hat{x}) \hat{d}^{\dagger} \hat{d},
\end{equation}
where $\hat p$ and $\hat x$ are the momentum operator and the position operator for the nuclei, $m_{\text{n}}$ is the mass of the nuclei, $\hat d$ and $\hat d^{\dagger}$ are the fermionic annihilation and creation operators corresponding to the ground state electronic orbital. Here we are considering a two-level nuclei: when the molecule is neutral, $\hat d^{\dagger} \hat d=0$, and the nuclear potential is
$U_0(\hat x)$; when the molecule is charged, $\hat d^{\dagger} \hat d=1$, and the potential is $U_1(x) = U_0(\hat x)+h(\hat x)$.

The system's Hamiltonian represents a two-level electronic system and therefore could be rewritten in the following matrix form: 
\begin{equation}
\begin{aligned}
    \hat H_S
&=\left(\begin{array}{cc} \hat h_0(x) &  \\&  \hat h_1(x)\end{array}\right)\\
&= \left(\begin{array}{cc}-\frac{\varepsilon^2}{2}\Delta+U_0(x) &  \\&  -\frac{\varepsilon^2}{2}\Delta+U_1(x)\end{array}\right).
\end{aligned}
\end{equation}
Here $\varepsilon$ is the ratio between the energy scale we are interested in and the macroscopic energy scale, which is a moderately small non-dimensionalized constant but fixed in the derivation, and it is also known as the semiclassical parameter \cite{cao_lindblad_2017}. 
The system wavefunction is
$$  \boldsymbol \psi(x) =   \left(\begin{array}{c}
    \psi_0(x)\\ \psi_1(x)
    \end{array}\right)\in \mathbb L^2(\mathbb R^n)\otimes \mathbb C^2,$$ where $\psi_0(x)$ is the nuclei wavefunction when the molecule is neutral (i.e. under potential $U_0(x)$), while $\psi_1(x)$ is the nuclei wavefunction when the molecule is charged (i.e. under potential $U_1(x)$). We also introduce the following Dirac notation, as it is widely used in literature:
    $$
    |\Psi\rangle = |\Psi_0\rangle|0\rangle + |\Psi_1\rangle |1\rangle,
    $$
    and in this notation, $\hat d$ and $\hat d^{\dagger}$ acts as $\hat d|1\rangle = |0\rangle, \hat d^{\dagger}|0\rangle = |1\rangle, \hat d|0\rangle = 0, \hat d^{\dagger}|1\rangle = 0$, and we have
\begin{equation} 
   \begin{aligned}
            \hat d|\Psi\rangle &= \hat d|\Psi_0\rangle|0\rangle + \hat d|\Psi_1\rangle |1\rangle = |\Psi_1\rangle |0\rangle\\
    \hat d^{\dagger}|\Psi\rangle& = \hat d^{\dagger}|\Psi_0\rangle|0\rangle + \hat d^{\dagger}|\Psi_1\rangle |1\rangle = |\Psi_0\rangle |1\rangle.
   \end{aligned}
    \label{eq:ddirac}
\end{equation}
In \cref{eq:ddirac}, when acting on $|\Psi\rangle\in\mathbb L^2(\mathbb R^n)\otimes \mathbb C^2$, $\hat d$ acts as  $\hat I\otimes \hat d$ (so is $\hat d^{\dagger}$).

The heat bath is formed by non-interacting metal orbitals $|E\rangle$ with energy $E\in [E_-,E_+]$, where $E_-$ and $E_+$ are the lower and upper bound of the spectrum of the metal continuum band.  As a result, the bath Hamiltonian is
\begin{equation}
    \hat{H}_B  =\int_{E_-}^{E_+} (E-\mu) \hat{c}_{E}^{\dagger} \hat{c}_E \mathrm d E.
\end{equation}
Here $\hat{c}_E$ and $\hat{c}_{E}^{\dagger}$  are the annihilation and creation operators for the metal electronic orbital $|E\rangle$, and $\mu$ is the chemical potential. For formal simplicity, $\mu$ is chosen to be zero unless otherwise specified, but all of our derivations naturally apply to any choice of $\mu$. 
For the sake of practical calculations,
the continuous band $[E_-,E_+]$ could be discretized using equidistant grid point \cite{huang2023efficient}:
\begin{equation}
    E_k = E_- + {(k-\frac{1}{2})}h_N,\quad  h_N = \frac{E_+-E_-}{N}.
\end{equation}
or using Gaussian quadrature \cite{shenvi2009nonadiabatic}. After discretization, the bath Hamiltonian becomes:
\begin{equation}
    \hat{H}_B  =\sum_{k=1}^N E_k\hat{c}_{k}^{\dagger} \hat{c}_k.
\label{eq:bath_discretize}
\end{equation}

For now, we will proceed our discussion using the discretized metal orbitals. However,
as we will show later,  after deriving the nMSSE model, we will consider the 
continuous band limit by letting the number of metal orbitals $N$ go to infinity, through which we recover the metal continuum and thus eliminate the quadrature error of continuous band discretization. 

The system-bath interaction $\hat H_{S-B}$, which describes the interaction between the molecule and metal continuum, is as follows:
\begin{equation}
     \lambda \hat{H}_{S-B} =\lambda\left(\int\mathrm d E\, V(E,\hat{x}) \hat{c}_E^{\dagger} \hat{d}+\bar{V}(E,\hat{x}) \hat{d}^{\dagger} \hat{c}_E\right).
\label{eq:system_bath}
\end{equation}
Here $V(E,x)$ describes the interaction strength between the molecule and the bath orbital with energy $E$ while the  nuclei is at $ x$, and $\bar V(E,x)$ is its complex conjugate. 
With the same discretization orbitals $\{E_k\}_{k=1}^N$ chosen as above, we have
\begin{equation}
   \lambda \hat{H}_{S-B} =\lambda\left(\sum_{k=1}^N V_k(\hat{x}) \hat{c}_k^{\dagger} \hat{d}+\bar{V}_k(\hat{x}) \hat{d}^{\dagger} \hat{c}_k\right)   \label{eq:system_bath_discretize}
\end{equation} 
where $V_k(\hat x) = \sqrt{h_N} V(E,x)$.
We refer to \cite{shenvi2009nonadiabatic} for the derivation of \cref{eq:system_bath_discretize}.
Without loss of generality, let us assume that $V_k(x)$ is real-valued.
 
\subsection{Bogoliubov transformation of the system-bath coupling}
\label{subsec:Bogoliubov}
For clarity in the derivation, we adopt the Bogoliubov transformations, which were introduced for studying weakly interacting $\text{He}^4$ superfluid \cite{bogoliubov1947theory} and solving Bardeen–Cooper–Schrieffer theory \cite{bogoljubov1958new} to understand superconductivity.  Let us define the following pairs of Bogoliubov operators:
\begin{equation}
\begin{aligned}
       \hat B_{1k} =\hat c_k+\hat c_k^{\dagger},& \quad 
    \hat B_{2k} =\mathrm i ( \hat c_k-\hat c_k^{\dagger}),\\ 
     \hat g_1 =\hat d+\hat d^{\dagger},& \quad  \hat g_2 =\mathrm i ( \hat d-\hat d^{\dagger}).
\end{aligned}
     \label{eq:bogliubovop}
\end{equation}
Note that $\hat B_{1k},\hat B_{2k},\hat g_1,\hat g_2$ are all Hermitian operators, and we have
\begin{equation}
 \begin{aligned}
        \hat c_k = \frac{\hat B_{1k}-\mathrm i\hat  B_{2k}}{2}&,\quad \hat c_k^{\dagger} = \frac{\hat B_{1k}+\mathrm i\hat  B_{2k}}{2}\\\hat d = \frac{\hat g_{1}-\mathrm i\hat  g_{2}}{2}&,\quad \hat d^{\dagger} = \frac{\hat g_{1}+\mathrm i\hat  g_{2}}{2}.
 \end{aligned}
\end{equation}
In Dirac's notation, according to \cref{eq:ddirac}, we have
\begin{equation}
\begin{aligned}
     \hat g_1|\Psi\rangle& = (\hat d+\hat d^{\dagger})(|\psi_0\rangle|0\rangle+|\psi_1\rangle|1\rangle)\\& = |\psi_0\rangle|1\rangle+|\psi_1\rangle|0\rangle, \\
    \hat g_2|\Psi\rangle& = \mathrm{i}(\hat d-\hat d^{\dagger})(|\psi_0\rangle|0\rangle+|\psi_1\rangle|1\rangle)\\& = -\mathrm{i}|\psi_0\rangle|1\rangle+\mathrm{i}|\psi_1\rangle|0\rangle.
\end{aligned}
\end{equation}
Therefore the Bogoliubov operators $\hat g_1,\hat g_2$ acts like Pauli matrices for the two-level system:
\begin{equation}
    \hat g_1 = \sigma_x=\left(
\begin{array}{cc}
0     &1  \\
1     & 0
\end{array}
\right),\quad \hat g_2 = -\sigma_y=\left(
\begin{array}{cc}
0     &\mathrm i \\
-\mathrm i     & 0
\end{array}
\right).
\label{eq:Pauli}
\end{equation}
Note that $[B_{ik},g_j]_+ = 0$, we can rewrite the coupling Hamiltonian as:
\begin{equation}
    \begin{aligned}
&\hat H_{S-B} \\&= \sum_{k=1}^N V_k\left(\frac{\hat B_{1k}+\mathrm i\hat  B_{2k}}{2}\frac{\hat g_{1}-\mathrm i\hat  g_{2}}{2} +\frac{\hat g_{1}+\mathrm i\hat  g_{2}}{2}\frac{\hat B_{1k}-\mathrm i\hat  B_{2k}}{2}\right)\\
&= \sum_{k=1}^N V_k\left(\frac{\hat g_{1}+\mathrm i\hat  g_{2}}{2}\frac{\hat B_{1k}-\mathrm i\hat  B_{2k}}{2}-\frac{\hat g_{1}-\mathrm i\hat  g_{2}}{2}\frac{\hat B_{1k}+\mathrm i\hat  B_{2k}}{2} \right)\\
&=\sum_{k=1}^N \frac{\mathrm{i}}{2}V_k(x)\hat g_{2}\hat B_{1k}-\frac{\mathrm{i}}{2}V_k(x)\hat g_{1}\hat B_{2k} .
\end{aligned}
\label{eq:H_s-b}
\end{equation}
In open quantum systems, the system-bath interaction is often written as a sum of multiplication of system Hermitian operator and bath Hermitian operator:
\begin{equation}
    \hat H_{S-B} = \sum_{i=1}^2\sum_{k=1}^N \hat S_{ik} \hat B_{ik}.
\end{equation}
Combined with \cref{eq:H_s-b}, we know that the system operators $\hat S_{ik}$ are
\begin{equation}
    \hat S_{1k} = \frac{\mathrm{i}}{2} V_k(x)\hat g_2,\quad \hat S_{2k} = -\frac{\mathrm{i}}{2} V_k(x)\hat g_1.
\end{equation}
\subsection{Time correlation function of bath operators and the memory kernel}
\label{subsec:timecorrelation}
The non-Markovian Stochastic Schr\"{o}dinger Equation (nMSSE) was formally derived in \cite{gaspard1999non} and was applied to simulate spin-boson systems. 
For a general open quantum system where the interaction between system and bath is described as $\hat H_{S-B} = \sum_{\alpha}\hat S_{\alpha}\hat B_{\alpha}$,
the effective nMSSE dynamics for the system is
\begin{equation}
\begin{aligned}
\mathrm{i}\varepsilon\partial_t|\Psi(t)\rangle &=  \hat{H}_{{S}}|\Psi(t)\rangle + \lambda \sum_\alpha \eta_\alpha(t) \hat{S}_\alpha|\Psi(t)\rangle \\
- & \mathrm{i} \frac{\lambda^2}{\varepsilon}  \int_0^t \mathrm{~d} \tau \sum_{\alpha, \beta} C_{\alpha, \beta}(\tau) \hat{S}_\alpha \mathrm{e}^{-\frac{\mathrm{i}}{\varepsilon} \hat{H}_{{S}} \tau} \hat{S}_\beta  |\Psi(t-\tau)\rangle.
\end{aligned}
\label{eq:generalsse}
\end{equation}
Recall that $\varepsilon$ is a small non-dimensionalized parameter.
Here $\eta_{\alpha}(t)$ is the complex-valued  Gaussian stochastic noise satisfying
\begin{equation}
    \begin{array}{c}   \mathbb{E}\left(\eta_{\alpha}(t)\right) = 0,\quad \mathbb{E}\left(\eta_{\alpha}(t)\eta_{\beta}(t')\right) = 0,\\ \mathbb{E}\left(\eta_{\alpha}^*(t)\eta_{\beta}(t')\right) =C_{\alpha,\beta}(t-t'),
    \end{array}
\end{equation}
and
 the memory kernel $C_{\alpha,\beta}(\tau)$ 
 is the time-correlation function of the bath operators:
\begin{equation}
C_{\alpha, \beta}\left(t-t^{\prime}\right)=\operatorname{tr}_B \left(\hat\rho_B^{\mathrm{eq}} \hat B_{\alpha}(t) \hat B_{\beta}\left(t^{\prime}\right)\right),
\label{eq:generalcorrelation}
\end{equation}
where  $\hat\rho_B^{\mathrm{eq} }=\frac{1}{Z_B}\exp(-\beta\hat H_B)= \frac{1}{Z_B}\exp(-\hat H_B/k_BT)$ is the density matrix of the bath, which is in thermal equilibrium of temperature $T$, $k_B$ is the Boltzmann constant, $Z_B = \operatorname{tr}(\exp(-\hat H_B/k_BT))$ and $\hat B_{\alpha}(t) = \exp{(\frac{\mathrm{i}}{\varepsilon}\hat H_Bt)}B_{\alpha}\exp{(-\frac{\mathrm{i}}{\varepsilon}\hat H_Bt)}$ is  the Heisenberg representation of $\hat B_{\alpha}$. 
In other words, the derivation of SSE model for Anderson-Holstein impurities boils down to calculating the time-correlation function \cref{eq:generalcorrelation} of the stochastic noise, which is also the memory kernel of the non-Markovian term.

Now let us derive an explicit expression for \cref{eq:generalcorrelation}. Recall that $N$ is the number of discrete metal bath orbitals. The natural basis for the Fock space $\mathcal H_B$ of the bath, under the occupation number representation, would be 
$$
\mathcal H_B = \operatorname{span}\bigl\{\left|b_1\right\rangle\cdots\left|b_N\right\rangle,\quad b_1,\cdots,b_N\in\{0,1\}\bigr\},
$$
or, in the binary representation,
$$
\mathcal H_B = \operatorname{span} \left\{\left|b\right\rangle |b=\sum b_k2^{k-1}, b\in\{0,1,\cdots,2^N-1\} \right\}.
$$
In this basis, the bath Hamitonian $\hat H_B$ is a $2^N\times 2^N$ diagonal matrix:
\begin{equation}
\begin{array}{c}
        \left(\hat H_B \right)_{bb'}=\delta_{bb'}\mathcal{E}_b,\quad\mathcal{E}_b= \sum_k b_kE_k, \\\text{for } b,b'\in\{0,1,\cdots,2^N-1\},\quad b_1,\cdots,b_N\in\{0,1\}.
\end{array} 
    \label{eq:bathhamiltonian}
\end{equation}
For future reference, let us call $b_k$ the $k$-th digit of $b$.
Now we are ready to calculate the correlation function $C_{ik,i'k'}\left(t-t^{\prime}\right)$. We have
\begin{equation}
\begin{aligned}
C_{ik,i'k'}\left(t-t^{\prime}\right)&=\operatorname{tr}_{B}\left(\frac{1}{Z_B} \mathrm{e}^{-\beta\hat H_B} \hat B_{ik}(t) \hat B_{i'k'}\left(t^{\prime}\right)\right) \\&= \frac{1}{Z_B}\sum_{b=0}^{2^N-1} \mathrm{e}^{-\beta \mathcal{E}_b}\langle b|\hat B_{ik}(t) \hat B_{i'k'}\left(t^{\prime}\right)|b \rangle,
\end{aligned}
\end{equation}
where $Z_B=\sum_{b=0}^{2^N-1} \mathrm{e}^{-\beta \mathcal{E}_{b}}$ is the normalizing factor.
Using the resolution of identity $\mathbb{I} = \sum_{b'} |b'\rangle\langle b'|$, one can rewrite the correlation function as
\begin{equation}
 \begin{aligned}
&   C_{ik,i'k'}\left(t-t^{\prime}\right)\\
& = \frac{1}{Z_B}
\sum_{b=0}^{2^N-1}\sum_{b'=0}^{2^N-1}\mathrm{e}^{-\beta \mathcal{E}_b}\langle b|B_{ik}(t)|b'\rangle\langle b'| B_{i'k'}\left(t^{\prime}\right)|b\rangle. 
 \end{aligned}
 \label{eq:bb'1}
\end{equation}
Recall that $\hat B_{1k} = \hat f_{1k}=\hat c_k+\hat c_k^{\dagger},  \hat B_{2k} = \hat f_{2k}=\mathrm i ( \hat c_k-\hat c_k^{\dagger})$, and note that
\begin{equation}
    \begin{aligned}
        \langle b|c_k(t)|b'\rangle &= \langle b|\mathrm{e}^{\frac{\mathrm{i}}{\varepsilon} \hat H_{B} t}c_k\mathrm{e}^{-\frac{\mathrm{i}}{\varepsilon}\hat H_{B} t}|b'\rangle \\&=\mathrm{e}^{\frac{\mathrm{i}}{\varepsilon}(\mathcal{E}_{b}-\mathcal{E}_{b'}) t}\langle b|c_k|b'\rangle
    ,
    \end{aligned}
    \label{eq:bb'2}
\end{equation}
Here we use $\mathrm{e}^{-\frac{\mathrm{i}}{\varepsilon}\hat H_Bt}|b\rangle = \mathrm{e}^{-\frac{\mathrm{i}}{\varepsilon}\mathcal{E}_bt}|b\rangle$ according to \cref{eq:bathhamiltonian}.
By definition of annihilation operators, $\langle b|c_k|b'\rangle$ is  nonzero if and only if  in the binary representation, $b_k=0$, $b'_k=1$, and except for the $k$-th digit, $b$ and $b'$ are the same.
Similarly,
\begin{equation}
\begin{aligned}
        \langle b|c_k^{\dagger}(t)|b'\rangle &= \langle b|\mathrm{e}^{\frac{\mathrm{i}}{\varepsilon} \hat H_{B} t}c_k^{\dagger}\mathrm{e}^{-\frac{\mathrm{i}}{\varepsilon} \hat H_{B} t}|b'\rangle \\&=
    \mathrm{e}^{\frac{\mathrm{i}}{\varepsilon}(\mathcal{E}_{b}-\mathcal{E}_{b'}) t}\langle b|c_k^{\dagger}|b'\rangle,
\end{aligned}
    \label{eq:bb'3}
\end{equation}
where $\langle b|c_k^{\dagger}|b'\rangle$ is by definition nonzero if and only if  
 in binary representation, $b_k=1$, $b'_k=0$, and except for the $k$-th digit, $b$ and $b'$ are the same.

In other words, given $b\in\{0,\cdots,2^N-1\}$, there is a unique $b'$ such that $\langle b|c_k(t)|b'\rangle$ ($\langle b|c_k^{\dagger}(t)|b'\rangle$) is nonzero if its $k$-th digit $b_k=1$ ($b_k=0$). This greatly simplifies the double sum in \cref{eq:bb'1}. We have
\begin{equation}
  \begin{aligned}
 & \sum_{b=0}^{2^N-1}\sum_{b'=0}^{2^N-1}\mathrm{e}^{-\beta \mathcal{E}_b}\langle b|c_k(t)|b'\rangle\langle b'| c_{k'}^{\dagger}\left(t^{\prime}\right)|b\rangle\\=&\delta_{kk'}\sum_{\{b |b_k= 0\}}\mathrm{e}^{-\beta \mathcal{E}_b} \mathrm{e}^{-\frac{\mathrm{i}}{\varepsilon} E_k (t-t')},
  \end{aligned}
    \label{eq:bb'4}
\end{equation}
\begin{equation}
\begin{aligned}
&\sum_{b=0}^{2^N-1}\sum_{b'=0}^{2^N-1}\mathrm{e}^{-\beta \mathcal{E}_b}\langle b|c_{k'}^{\dagger}(t)|b'\rangle\langle b'| c_k\left(t^{\prime}\right)|b\rangle\\=&\delta_{kk'}\sum_{\{b |b_k= 1\}}\mathrm{e}^{-\beta \mathcal{E}_b} \mathrm{e}^{\frac{\mathrm{i}}{\varepsilon} E_k (t-t')},
\end{aligned}
    \label{eq:bb'5}
\end{equation}
Let 
$$
\begin{aligned}
C_{k}^+(\tau)  &=\frac{1}{Z_B} \sum_{\{b |b_k= 0\}}\mathrm{e}^{-\beta \mathcal{E}_b} \mathrm{e}^{-\frac{\mathrm{i}}{\varepsilon} E_k \tau},\\ C_{k}^-(\tau)  & = \frac{1}{Z_B}\sum_{\{b |b_k= 1\}}\mathrm{e}^{-\beta \mathcal{E}_b} \mathrm{e}^{\frac{\mathrm{i}}{\varepsilon} E_k \tau},
\end{aligned}
$$
Combining \cref{eq:bb'1,eq:bb'2,eq:bb'3}, we have\begin{equation}
    C_{ik,i'k'}(\tau ) = \delta_{kk'}C_{ii',k}(\tau),
\end{equation}
where
\begin{equation}
    \begin{aligned}
&        C_{11,k}(\tau) = C_{k}^+(\tau) + C_{k}^-(\tau),\\&
 C_{22,k}(\tau) =  C_{k}^+(\tau) + C_{k}^-(\tau), \\
 &      C_{12,k}(\tau) = -\mathrm{i}C_{k}^+(\tau) +\mathrm{i} C_{k}^-(\tau),\\&
    C_{21,k}(\tau) =\mathrm{i} C_{k}^+(\tau) - \mathrm{i}C_{k}^-(\tau) . 
    \end{aligned}
    \label{eq:crelations}
\end{equation}
What's left to do is to calculate out $C_k^{\pm}(\tau)$. Note that 
$$
 C_{k}^+(\tau) + C_{k}^-(-\tau)  =\frac{1}{Z_B}\sum_b \mathrm{e}^{-\beta \mathcal{E}_b} \mathrm{e}^{-\frac{\mathrm{i}}{\varepsilon} E_k \tau} =  \mathrm{e}^{-\frac{\mathrm{i}}{\varepsilon} E_k \tau},
$$
$$
\frac{ C_{k}^+(\tau)}{ C_{k}^-(-\tau)} = \frac{\sum_{\{b |b_k= 0\}}\mathrm{e}^{-\beta \mathcal{E}_b} \mathrm{e}^{-\frac{\mathrm{i}}{\varepsilon} E_k (t-t')}}{\sum_{\{b |b_k= 1\}}\mathrm{e}^{-\beta \mathcal{E}_b} \mathrm{e}^{-\frac{\mathrm{i}}{\varepsilon} E_k (t-t')}} = \mathrm{e}^{\beta E_k},
$$
therefore we have
\begin{equation}
    C_{k}^+(\tau) =\frac{\exp{(-\frac{\mathrm{i}}{\varepsilon}E_k\tau)}}{1+\exp{(-\beta E_k)}} ,\quad C_{k}^-(\tau) = \frac{\exp{(\frac{\mathrm{i}}{\varepsilon} E_k\tau)}}{1+\exp{(\beta E_k)}}.
    \label{eq:defCkpm}
\end{equation}
\subsection{Towards the nMSSE model}
\label{subsec:nMSSE}
Now we are ready to write down the nMSSE for the Anderson-Holstein model. Let us derive the formula in the spatial representation.
The wavefunction has two components:
\begin{equation}
    \Psi(x,t) = \left(
    \begin{array}{c}
        \psi_0(x,t)\\
        \psi_1(x,t)
    \end{array}
    \right).
\end{equation}
The first term on the right hand side of \cref{eq:generalsse} is the system Hamiltonian itself:
\begin{equation}
\begin{aligned}
    &\hat H_S\left(\begin{array}{c}\psi_0(x,t)\\ \psi_1(x,t)\end{array}\right) \\=& \left(\begin{array}{cc}
    -\frac{\varepsilon^2}{2}\Delta+U_0(x)     &  \\
         &   -\frac{\varepsilon^2}{2}\Delta+U_1(x) 
    \end{array}
    \right)\left(\begin{array}{c}\psi_0(x,t)\\ \psi_1(x,t)\end{array}\right).
\end{aligned}
    \label{eq:firstterm}
\end{equation}
The second term is the stochastic noise, and could be rewritten as
\begin{equation}
    \begin{aligned}
   &  \lambda \sum_{ik}\eta_{ik}(t) \hat{S}_{ik}\Psi(x,t)\\ &
   \begin{aligned}
       = \lambda \sum_{k}&
   \left(\frac{\mathrm{i}}{2} V_k(x)\eta_{1k}(t)\left(
    \begin{array}{c}
        \mathrm i\psi_1\\
        -\mathrm i\psi_0
    \end{array}
    \right)\right.\\ -& \left.\frac{\mathrm{i}}{2} V_k(x)\eta_{2k}(t)\left(
    \begin{array}{c}
        \psi_1\\
        \psi_0
    \end{array}
    \right) \right)
   \end{aligned}
     \\
     =&  -\frac{\mathrm{i} \lambda}{2}\sum_{k=1}^N V_k(x)\left(\eta_{2k}(t) \sigma_x+\eta_{1k}(t) \sigma_y\right)\left(\begin{array}{l}
\psi_0(x, t) \\
\psi_1(x, t)
\end{array}\right),
\end{aligned}
\label{eq:stochasticnoisefinal}
\end{equation}
and  the stochastic noise $\eta_{ik}$ satisfies that:
\begin{equation}
\begin{array}{c}
\mathbb{E}(\eta_{ik}(t))=0, \quad \mathbb{E}(\eta_{ik}(t)\eta_{i'k'}(t'))=0,
\\
\mathbb{E}\left(
\eta_{ik}^{*}(t) \eta_{i'k'}\left(t^{\prime}\right)\right) = {\delta_{kk'}}
C_{ii',k}(t-t'),\quad i,i'=1,2. 
\end{array}
\label{eq:noisediscrete}
\end{equation}

The non-Markovian damping term is
$$
-\mathrm i \frac{\lambda^{2} }{\varepsilon}\int_{0}^{t} \mathrm d \tau \sum_{ik,i'k'} C_{ik,i'k'}(\tau) \hat S_{ik} \mathrm e^{-\mathrm i \hat H_{s} \tau} \hat S_{i'k'} \psi(x,t-\tau).$$
Note that
\begin{equation}
    \mathrm e^{-\frac{\mathrm{i}}{\varepsilon} \hat H_{s} \tau} = \left(\begin{array}{cc}   \mathrm e^{-\frac{\mathrm{i}}{\varepsilon}\hat h_0\tau} &  \\&  \mathrm e^{-\frac{\mathrm{i}}{\varepsilon}\hat h_1\tau}\end{array}\right)
\end{equation}
With a straightforward calculation, we have
\begin{equation}
    \begin{aligned}
& \sum_{i,j=1}^2\sum_{k=1}^N C_{ij,k}(\tau) \hat S_{ik} e^{-\frac{\mathrm{i}}{\varepsilon} \hat H_{s} \tau} \hat S_{jk} \psi(x,t-\tau) \\
 =-&\sum_{k=1}^N
 \left(\begin{array}{c}  C_{k}^-(\tau)V_k(x) \mathrm e^{-\frac{\mathrm{i}}{\varepsilon}\hat h_1\tau } {V}_k(x) \psi_0(x,t-\tau) \\   C_{k}^+(\tau) V_k(x) \mathrm e^{-\frac{\mathrm{i}}{\varepsilon}\hat h_0\tau}V_k(x)\psi_1(x,t-\tau)\end{array}\right).
\end{aligned}
\label{eq:nonMarkovianterm}
\end{equation}
Here $\mathrm{Diag}\left(\begin{array}{c} a\\b\end{array}\right)$ means $\left(\begin{array}{cc} a& 0\\0 &b\end{array}\right)$.
Note that operators $\mathrm{e}^{-\mathrm i \hat h_{0,1}\tau}$ and $V_k$ typically does not commute.

Combining \cref{eq:generalsse,eq:firstterm,eq:stochasticnoisefinal,eq:nonMarkovianterm}, we arrive at the discretized nMSSE model for AH impurities:
\begin{equation}
\begin{aligned}
\mathrm{i}\varepsilon& \frac{\partial}{\partial t}\left(\begin{array}{c}\psi_0(x,t)\\ \psi_1(x,t)\end{array}\right)=\left(\begin{array}{c}\hat{h}_0\psi_0(x,t)\\ \hat{h}_1\psi_1(x,t)\end{array}\right)\\
-&\frac{\mathrm{i} \lambda}{2}\sum_{k=1}^N V_k(x)\left(\eta_{2k}(t) \sigma_x+\eta_{1k}(t) \sigma_y\right)\left(\begin{array}{l}
\psi_0(x,t) \\
\psi_1(x,t)
\end{array}\right) \\ +&{\mathrm i }\frac{\lambda^{2} }{\varepsilon}
\int_0^t\mathrm{d}\tau
 \left(\begin{array}{c} \sum_{k} C_{k}^-(\tau)V_k \mathrm e^{-\frac{\mathrm{i}}{\varepsilon}\hat h_1\tau } {V}_k \psi_0(x,t-\tau) \\  \sum_{k} C_{k}^+(\tau) V_k \mathrm e^{-\frac{\mathrm{i}}{\varepsilon}\hat h_0\tau}V_k \psi_1(x,t-\tau)\end{array}\right).
\end{aligned}
\label{eq:nMSSEmetal}
\end{equation}
where $\hat h_{0,1} = -\frac{\varepsilon^2}{2}\Delta+U_{0,1}(x)$, and $\sigma_{x,y}$ is the Pauli matrices. Complex-valued Gaussian noise $\eta_{ik}(t)$  is defined in \cref{eq:noisediscrete}, and the memory kernel  $C_{k}^{\pm}(\tau)$ is defined in \cref{eq:defCkpm}.

\subsection{Wide band limit and continuous band limit}
\label{subsec:widebandlimit}
In the wide band limit, the system-bath coupling $V(E,x)$ (or, in the discrete setting, $V_k(x)$) is considered to be independent of the metal spectrum $E$ (or the discrete band index $k$), i.e. $V(E,x)=V(x)$ for any $E\in [E_-,E_+]$, and $V_k(x) = \sqrt{h_N}V(x)$. 
Let us define the total noise $\xi^{(N)}_i(t)$ 
$$
\xi_i^{(N)}(t) = \sqrt{h_N}\sum_{k=1}^N\eta_{ik}(t).
$$
Recalling \cref{eq:noisediscrete}, we have
\begin{equation}
 \begin{aligned}
 & \mathbb{E}(\xi_i^{(N)}(t))=0, \quad \mathbb{E}(\xi_i^{(N)}(t)\xi_{i'}(t'))=0 ,\\
    & \mathbb{E}(\xi^{(N)*}_i(t)\xi^{(N)}_{i'}(t'))\\ =& h_N\sum_{kk'}\mathbb{E}\left(
\eta_{ik}^{*}(t) \eta_{i'k'}\left(t^{\prime}\right)\right) = h_N\sum_{k=1}^N C_{ii',k}(t-t'),
 \end{aligned}
\label{eq:correlationbigN}
\end{equation}
For $i=1$, $i'=1$, we have
$$
\begin{aligned}
    \mathbb{E}(\xi^{(N)*}_1(t)\xi^{(N)}_{1}(t')) &= h_N\sum_{k=1}^N C_{11,k}(t-t')\\&=h_N\sum_{k=1}^N\left( C_{k}^+(t-t') + C_{k}^-(t-t')\right).
\end{aligned}
$$
Taking the continuous band limit, i.e. let $N\rightarrow\infty$, we have
$$
\begin{aligned}
    h_N\sum_{k=1}^N C_k^{\pm}(\tau) &= h_N\sum_{k=1}^N\frac{\exp{(\mp\frac{\mathrm{i}}{\varepsilon}E_k\tau)}}{1+\exp{(\mp\beta E_k)}}
    \\
    &\rightarrow \int_{E_-}^{E_+} \frac{\exp{(\mp\frac{\mathrm{i}}{\varepsilon}E\tau)}}{1+\exp{(\mp\beta E)}}\mathrm{d}E=:c^{\pm}(\tau).
\end{aligned}
$$
For $(i,i')=(1,2), (2,1), (2,2)$, things are similar. Therefore when $N\rightarrow\infty$, $\xi_i^{(N)}(t)$ converges to $\xi_i(t)$, which satisfies
\begin{equation}
\begin{aligned}
   \mathbb{E}(\xi_i(t))=0, \quad &\mathbb{E}(\xi_i(t)\xi_{i'}(t'))=0 ,\quad i=1,2,\\
 \mathbb{E}(\xi^{*}_1(t)\xi_{1}(t'))&=  c^+(t-t') + c^-(t-t'),\\
\mathbb{E}(\xi^{*}_2(t)\xi_{2}(t'))&=  c^+(t-t') + c^-(t-t'), \\
  \mathbb{E}(\xi^{*}_1(t)\xi_{2}(t'))&= -\mathrm{i}c^+(t-t') +\mathrm{i} c^-(t-t'),\\
   \mathbb{E}(\xi^{*}_2(t)\xi_{1}(t')) &=\mathrm{i} c^+(t-t') - \mathrm{i}c^-(t-t').  
\end{aligned}
\label{eq:correlationcontinuous}
\end{equation}
Now the noise term \cref{eq:stochasticnoisefinal} becomes:
\begin{equation}
    \begin{aligned}
   & -\frac{\mathrm{i} \lambda}{2}\sum_{k=1}^N V_k(x)\left(\eta_{2k}(t) \sigma_x+\eta_{1k}(t) \sigma_y\right)\left(\begin{array}{l}
\psi_0(x, t) \\
\psi_1(x, t)
\end{array}\right) \\&\xrightarrow{N\rightarrow\infty} -\frac{\mathrm{i}\lambda}{2}V(x)\left(\xi_2(t) \sigma_x+\xi_1(t) \sigma_y\right)\left(\begin{array}{l}
\psi_0(x, t) \\
\psi_1(x, t)
\end{array}\right).
\end{aligned}
\label{eq:noisewidecont}
\end{equation}
In the wide band limit and the continuous band limit, we also have
$$
\begin{aligned}
   & \sum_{k=1}^N C_{k}^{\pm}(\tau)V_k(x) \mathrm e^{-\frac{\mathrm{i}}{\varepsilon}\hat h_{0,1}\tau } {V}_k(x)\\&=\sum_{k=1}^N h_N C_{k}^{\pm}(\tau)V(x) \mathrm e^{-\frac{\mathrm{i}}{\varepsilon}\hat h_{0,1}\tau } {V}(x) \\&\rightarrow  c^{\pm}(\tau)V(x) \mathrm e^{-\frac{\mathrm{i}}{\varepsilon}\hat h_{0,1}\tau } {V}(x).
\end{aligned}
$$
Therefore \cref{eq:nonMarkovianterm} becomes
\begin{equation}
-
 \left(\begin{array}{c}  c^-(\tau)V(x) \mathrm e^{-\frac{\mathrm{i}}{\varepsilon}\hat h_1\tau } {V}(x)\psi_0(x,t-\tau)   \\  c^+(\tau) V(x) \mathrm e^{-\frac{\mathrm{i}}{\varepsilon}\hat h_0\tau}V(x)\psi_1(x,t-\tau)\end{array}\right). \label{eq:nonmarkovianwidecont}
\end{equation}
Replacing the noise and non-Markovian terms in \cref{eq:nMSSEmetal} with \cref{eq:noisewidecont,eq:nonmarkovianwidecont}
, we arrive at the nMSSE model in the wide and continuous band limit:

\begin{equation}
\begin{aligned}
\mathrm{i}\varepsilon \frac{\partial}{\partial t}&\left(\begin{array}{c}\psi_0(x,t)\\ \psi_1(x,t)\end{array}\right)=\left(\begin{array}{c}\hat{h}_0\psi_0(x,t)\\ \hat{h}_1\psi_1(x,t)\end{array}\right)\\
        -&\frac{\mathrm{i} \lambda V(x)}{2}\left(\xi_2(t) \sigma_x+\xi_1(t) \sigma_y\right)\left(\begin{array}{l}
\psi_0(x, t) \\
\psi_1(x, t)
\end{array}\right)  \\ +&\mathrm i \frac{\lambda^{2} }{\varepsilon}
 \int_{0}^{t} \mathrm d \tau 
 \left(
 \begin{array}{c}
 c^-(\tau) V(x) e^{-\frac{\mathrm{i}}{\varepsilon}\hat h_1\tau} V(x)  \psi_0(x,t-\tau)      \\
  c^+(\tau) V(x)\mathrm e^{-\frac{\mathrm{i}}{\varepsilon}\hat h_0\tau}  V(x)\psi_1(x,t-\tau)
 \end{array}
 \right).
\end{aligned}
\label{eq:nMSSEmetal_wide}
\end{equation}

The wide band limit still retained interesting physics since the separation of variables still preserves the spatial-inhomogeneity of the stochastic noise.

In the infinite temperature limit, i.e. $\beta=0$, if $E_{\pm} = \pm\infty$, then the correlation function becomes
\begin{equation}
c^{\pm}(\tau) = \int_{-\infty}^{+\infty}\exp\left(\mp \frac{\mathrm i}{\varepsilon}E\tau\right)\mathrm{d}\tau=\pm 2\pi\varepsilon\delta(\tau).
\end{equation}
\section{From SSE to QME}
\label{sec:SSE2QME}
Quantum master equations (QME) are often used as semi-empirical models when studying open quantum systems such as metal surfaces \cite{dou_surface_2015,dou2016broadened,dou2017born}.  Here we show that QME models of the Anderson-Holstein impurities are a second-step approximation to the SSE models in the interaction picture, and the well-known Redfield equation could be achieved with a further Markovian approximation.
\subsection{Analytic derivation}
\label{subsec:derivation}
In \cref{eq:nMSSEmetal_wide},   we could think of the nMSSE effective Hamiltonian being $\hat H_S|\Psi\rangle + \hat H_{\mathrm{int}}|\Psi\rangle$, where
$
\hat H_{\mathrm{int}}
$ describes the effective interaction between the system and the bath:
\begin{equation}
\begin{aligned}
&\hat H_{\mathrm{int}}\left(\begin{array}{l}
\psi_0(x, t) \\
\psi_1(x, t)
\end{array}\right) \\=&  -\frac{\mathrm{i} \lambda V(x)}{2}\left(\xi_2(t) \sigma_x+\xi_1(t) \sigma_y\right)\left(\begin{array}{l}
\psi_0(x, t) \\
\psi_1(x, t)
\end{array}\right)  \\& +{\mathrm i }\frac{\lambda^{2} }{\varepsilon}
 \int_{0}^{t} \mathrm d \tau
\left(
 \begin{array}{c}
 c^-(\tau)V(x)\mathrm e^{-\frac{\mathrm i}{\varepsilon}\hat h_1\tau}   V(x) \psi_0(x,t-\tau)   \\
       c^+(\tau)V(x)\mathrm e^{-\frac{\mathrm i}{\varepsilon}\hat h_0\tau} V(x)\psi_1(x,t-\tau)
 \end{array}
 \right)
 \\
=& \left( \lambda \hat H^{(1)}_{\mathrm{int}}(t) + \frac{\lambda^{2} }{\varepsilon} \hat H^{(2)}_{\mathrm{int}}(t)\right)\left(\begin{array}{c}
\psi_0(x,t)\\ \psi_1(x,t))\end{array}\right).
\end{aligned}
\label{eq:Hint}
\end{equation}
Here we define $\hat H^{(1)}_{\mathrm{int}}(t)$ and $ \hat H^{(2)}_{\mathrm{int}}(t)$ based on the order of $\lambda$.

Let $\Psi_I(x,t)$ be the wavefunction in the interaction picture. Then $\Psi_I(x,t)$ satisfies that:
\begin{equation}
   \mathrm{i} \varepsilon\frac{\partial}{\partial t}\Psi_I(x,t)
   = \hat H_{\mathrm{int},I}(t)\Psi_I(x,t),
   \label{eq:interactionpicture}
\end{equation}
where $\hat H_{\mathrm{int},I}(t)$ is $\hat H_{\mathrm{int}}(t)$ in the interation picture:
\begin{equation}
    \hat H_{\mathrm{int},I}(t) = \mathrm{e}^{\frac{\mathrm{i}}{\varepsilon}\hat H_S t}\hat H_{\mathrm{int}}(t)\mathrm{e}^{-\frac{\mathrm{i}}{\varepsilon}\hat H_S t}.
\end{equation}
Note that we have the expansion $\hat H_{\mathrm{int},I}(t) = \lambda \hat H^{(1)}_{\mathrm{int},I}(t) + \frac{\lambda^{2} }{\varepsilon} \hat H^{(2)}_{\mathrm{int},I}(t)$. 

The quantum master equation describes the evolution of the expectation value of the density operator $\hat \rho_I = |\Psi_I\rangle\langle \Psi_I|$.
If  the wavefunction 
 $|\Psi_I\rangle$ has the  following asymptotic expansion:
$$
|\Psi_I\rangle = |\Psi^{(0)}_I\rangle + \frac{\lambda}{\varepsilon}|\Psi^{(1)}_I\rangle + \left(\frac{\lambda}{\varepsilon}\right)^2|\Psi^{(2)}_I\rangle + O\left(\left(\frac{\lambda}{\varepsilon}\right)^3\right),
$$
then the expectation of the density matrix becomes
$$
\begin{aligned}
  \mathbb{E}  \hat \rho_I &= \mathbb{E}\hat \rho^{(0)}_I+ \frac{\lambda}{\varepsilon} \mathbb{E}\hat \rho^{(1)}_I +\left(\frac{\lambda}{\varepsilon}\right)^2 \mathbb{E}\hat \rho^{(2)}_I + O\left(\left(\frac{\lambda}{\varepsilon}\right)^3\right),
    \\
    \hat \rho^{(0)}_I& =|\Psi_I^{(0)}\rangle\langle \Psi_I^{(0)}| , \quad \hat\rho^{(1)}_I =|\Psi_I^{(0)}\rangle\langle \Psi_I^{(1)}| + |\Psi_I^{(1)}\rangle\langle \Psi_I^{(0)}|, \\
    \hat\rho^{(2)}_I &= |\Psi_I^{(0)}\rangle\langle \Psi_I^{(2)}| + |\Psi_I^{(2)}\rangle\langle \Psi_I^{(0)}|+|\Psi_I^{(1)}\rangle\langle \Psi_I^{(1)}| . 
\end{aligned}
$$
In the current work, $\varepsilon$ is fixed, and considered to be $O(1)$.

Now we want to find $|\Psi^{(0)}_I\rangle , |\Psi^{(1)}_I\rangle , |\Psi^{(2)}_I\rangle$. 
Intergrating \cref{eq:interactionpicture}, and then substituting into itself, recalling that $ \hat H_{\mathrm{int},I}(t) = \lambda \hat H^{(1)}_{\mathrm{int},I}(t) + \frac{\lambda^2}{\varepsilon} \hat H^{(2)}_{\mathrm{int},I}(t)$ we have
\begin{equation}
\begin{aligned}
  &  \Psi_I(x,t)\\ & = \Psi_I(x,0) -\frac{\mathrm{i}}{\varepsilon} \int_{0}^t\hat H_{\mathrm{int},I}(t_1)\Psi_I(x,t_1)\mathrm{d}t_1\\
    & = \Psi_I(x,0) -\frac{\mathrm{i}}{\varepsilon} \left(\int_{0}^t\hat H_{\mathrm{int},I}(t_1)\mathrm{d}t_1\right)\Psi_I(x,0) \\&+ \left(-\frac{\mathrm{i}}{\varepsilon}\right)^2 \int_{0}^t\mathrm{d}t_1 \int_{0}^{t_1}\mathrm{d}t_2\hat H_{\mathrm{int},I}(t_1)H_{\mathrm{int},I}(t_2)\Psi_I(x,t_2)\\
    & =  \Psi_I(x,0) -\frac{\mathrm{i}}{\varepsilon}\lambda \left(\int_{0}^t\hat H_{\mathrm{int},I}^{(1)}(t_1)\mathrm{d}t_1\right)\Psi_I(x,0)\\&-\frac{\mathrm{i}}{\varepsilon}\frac{\lambda^{2} }{\varepsilon}\left(\int_{0}^t\hat H_{\mathrm{int},I}^{(2)}(t_1)\mathrm{d}t_1\right)\Psi_I(x,0)\\
    & + \left(-\frac{\mathrm{i}}{\varepsilon}\right)^2 \lambda^2\int_{0}^t\mathrm{d}t_1 \int_{0}^{t_1}\mathrm{d}t_2\hat H^{(1)}_{\mathrm{int},I}(t_1)H^{(1)}_{\mathrm{int},I}(t_2)\Psi_I^{(0)}(x,0)\\&+O\left(\left(\frac{\lambda}{\varepsilon}\right)^3\right).
\end{aligned}
\end{equation}
Therefore we have
\begin{equation}
    \begin{aligned}
       \Psi^{(0)}_I(x,t) &=   \Psi_I(x,0),\\
       \Psi^{(1)}_I(x,t) &=  -{\mathrm{i}}\left(\int_{0}^t\hat H_{\mathrm{int},I}^{(1)}(t_1)\mathrm{d}t_1\right)\Psi_I(x,0),\\
       \Psi^{(2)}_I(x,t) &=-{\mathrm{i}}\left(\int_{0}^t\hat H_{\mathrm{int},I}^{(2)}(t_1)\mathrm{d}t_1\right)\Psi_I(x,0)  \\&- \int_{0}^t\mathrm{d}t_1 \int_{0}^{t_1}\mathrm{d}t_2\hat H^{(1)}_{\mathrm{int},I}(t_1)H^{(1)}_{\mathrm{int},I}(t_2)\Psi_I(x,0).
    \end{aligned}
    \label{eq:psi_expansion}
\end{equation}
Note that $\mathbb E\Psi_I^{(1)}(x,t) = 0$ since the noise term has zero mean, therefore
\begin{equation}
    \begin{aligned}
\mathbb E\hat \rho_I^{(0)} &=\mathbb E\left(|\Psi_I^{(0)}\rangle\langle \Psi_I^{(0)}|\right) = |\Psi_I^{(0)}\rangle\langle \Psi_I^{(0)}|.\\
    \mathbb E\hat \rho_I^{(1)} &= \mathbb E\left(|\Psi_I^{(0)}\rangle\langle \Psi_I^{(1)}| + |\Psi_I^{(1)}\rangle\langle \Psi_I^{(0)}|\right) \\&=|\Psi_I^{(0)}\rangle \mathbb E\left(\langle \Psi_I^{(1)}|\right) + \mathbb E\left(|\Psi_I^{(1)}\rangle\right)\langle \Psi_I^{(0)}| = 0. \\
    \mathbb E\hat\rho_I^{(2)} &= \mathbb E\left(|\Psi_I^{(0)}\rangle  \langle \Psi_I^{(2)}| + |\Psi_I^{(2)}\rangle\langle \Psi_I^{(0)}|+|\Psi_I^{(1)}\rangle\langle \Psi_I^{(1)}|\right)\\&
    \begin{aligned}
        &=|\Psi_I^{(0)}\rangle
    \mathbb{E}\left(\langle \Psi_I^{(2)}|\right) + \mathbb{E}\left(|\Psi_I^{(2)}\rangle\right)\langle \Psi_I^{(0)}|\\&+\mathbb E\left(|\Psi_I^{(1)}\rangle\langle \Psi_I^{(1)}|\right).  
    \end{aligned}
    \end{aligned}
\label{eq:expectationsdetail}
\end{equation}
Let $\rho_I(t) = \mathbb E \hat \rho_I(t)$. With \cref{eq:expectationsdetail}
, we can differentiate $\rho_I(t)$ with respect to $t$ and truncate terms that are higher than second order in $\lambda$, and then obtain the non-Markivian quantum master equation (nMQME):

\begin{equation}
    \begin{aligned}
  & \begin{aligned}
      \frac{\mathrm{d} \rho_I(x,x^{\prime},t)}{\mathrm{d} t}=  
  &  \frac{\lambda^2}{\varepsilon^2} 
    \left(
        \int_0^t\mathbf c_h(x,\tau)\rho_I(x,x^{\prime},t-\tau)\mathrm{d}\tau
 \right.\\  
 \int_0^t \mathbf c(x,x^{\prime},\tau)&\left.\rho_I^{(d)}(x,x^{\prime},t-\tau)\mathrm{d}\tau
+ \mathrm{h.c.} \right) + O\left(\left(\frac{\lambda}{\varepsilon}\right)^3 \right)
    \end{aligned}\\
 & \rho_I(0)=\rho_0,
\end{aligned}
 \label{eq:QME}
\end{equation}
where $\mathrm{h.c.}$ represents the Hermitian conjugate of the preceding terms, and 
$$
\begin{aligned}
&\mathbf c(x,x^{\prime},\tau) = \left(
 \begin{array}{cc}
  c^+(\tau)V(x)V(x^{\prime})      &  \\
      & c^-(\tau)V(x)V(x^{\prime})
 \end{array}
 \right),\\ 
 &\mathbf c_h(x,\tau)\\& =\left(
 \begin{array}{cc}
  c^-(\tau)V(x)\mathrm e^{-\mathrm i\hat h_1\tau}V(x)      &  \\
      & c^+(\tau)V(x)\mathrm e^{-\mathrm i\hat h_0\tau} V(x)
 \end{array}
 \right), \\
 &\rho_I^{(d)}(x,x^{\prime},\tau) = \left(
 \begin{array}{cc}
  \rho_{I,11}(x,x^{\prime},\tau)      &  \\
      & \rho_{I,00}(x,x^{\prime},\tau)
 \end{array}
 \right).
\end{aligned}
$$
For the detailed derivation of \cref{eq:QME}, which reduces to the further simplification of \cref{eq:expectationsdetail} and its derivative, please see \cref{appendix:QMEderivation}.

More generally, the above master equation could be derived starting with any initial time $t_0$ other than time $t_0=0$. In that case, the quantum master equation becomes
\begin{equation}
    \begin{aligned}
   \frac{\mathrm{d} \rho_I(x,x^{\prime},t)}{\mathrm{d} t} 
    & = \frac{\lambda^2}{\varepsilon^2} 
    \left(
        \int_0^{t-t_0}\mathbf c_h(x,\tau)\rho_I(x,x^{\prime},t-\tau)\mathrm{d}\tau \right.\\
+ 
 \int_0^{t-t_0}& \left. \mathbf c(x,x^{\prime},\tau)\rho_I^{(d)}(x,x^{\prime},t-\tau)\mathrm{d}\tau
 \right)+ \mathrm{h.c.},\\
 \rho_I(t_0)&=\rho_0.
\end{aligned}
 \label{eq:QME_t0}
\end{equation}

\begin{widetext}
In Schr\"{o}dinger picture, it becomes
\begin{equation}
    \begin{aligned}
       \frac{\mathrm{d}\rho(x,x^{\prime},t)}{\mathrm{d}t}&=-\mathrm{i}\left(\hat{H}_s(x)-\hat{H}_s(x^{\prime})\right)\rho(x,x^{\prime},t)
     \\&\begin{aligned}
         +\frac{\lambda^2}{\varepsilon^2}
        \mathrm{e}^{-\mathrm{i}\hat{H}_S(x)(t-t_0)}\mathrm{e}^{\mathrm{i}\hat{H}_S(x^{\prime})(t-t_0)}\times&\left(\int_{t_0}^{t}\mathrm{d}\tau\left(\mathbf c_h(x,t-\tau)\mathrm{e}^{\mathrm{i}\hat{H}_S(x)(\tau-t_0)}\mathrm{e}^{-\mathrm{i}\hat{H}_S(x^{\prime})(\tau-t_0)}\rho(x,x^{\prime},\tau)\right)\right.   \\+\int_{t_0}^{t}\mathrm{d}\tau & \left(\mathbf{c}(x,x^{\prime},t-\tau)\sigma_x\mathrm{e}^{\mathrm{i}\hat{H}_S(x)(\tau-t_0)}
        \mathrm{e}^{-\mathrm{i}\hat{H}_S(x^{\prime})(\tau-t_0)}
        \rho(x,x^{\prime},\tau)\sigma_x+\mathrm{h.c.}\right),
     \end{aligned}
      \\ 
      \rho(t_0)&=\rho_I(t_0)=\rho_0.
    \end{aligned}
\end{equation}
\end{widetext}
Here $\sigma_x = \left(
\begin{array}{cc}
0     &1  \\
1     & 0
\end{array}
\right)$ as mentioned in \cref{eq:Pauli}.

\subsection{Born-Markov approximation, finite-history Quantum Master Equation (FH-QME), Redfield equation, and their corresponding SSE}
\label{subsec:manyqme}

The non-Markovian quantum master equations \cref{eq:QME_t0} are mathematically complicated and computationally expansive. The widely-used Born-Markov approximation could be performed under the conditions that (1) the interaction between the system and the bath is weak, (2) the correlation time of the bath is much shorter than the characteristic time of the system.
With the basic assumption $\rho_I(t-\tau)\approx \rho_I(t)$, the nMQME is reduced to 
\begin{equation}
    \begin{aligned}
   \frac{\mathrm{d} \rho_I(x,x^{\prime},t)}{\mathrm{d} t} 
    & = \frac{\lambda^2}{\varepsilon^2} 
    \left(
        \int_0^{t-t_0}\mathbf c_h(x,\tau)\mathrm{d}\tau\rho_I(x,x^{\prime},t)\right.\\
 + 
 \int_0^{t-t_0} &\left.\mathbf c(x,x^{\prime},\tau)\mathrm{d}\tau\rho_I^{(d)}(x,x^{\prime},t)
 \right)+ h.c.,\\
 \rho_I(t_0)&=\rho_0.
\end{aligned}
 \label{eq:QME_markov}
\end{equation}
If $t_0$ is set to be $0$, we refer to \cref{eq:QME_markov} as the finite-history QME (FH-QME). 
 The corresponding SSE of FH-QME in the interaction picture could be easily obtained by modifying \cref{eq:Hint} with the approximation $\psi_{0,I}(x,t-\tau) \approx \psi_{0,I}(x,t)$ and $\psi_{1,I}(x,t-\tau) \approx \psi_{1,I}(x,t)$. 
\begin{equation}
\begin{aligned}
&\mathrm{i}\varepsilon \frac{\partial}{\partial t} \left(\begin{array}{c}\psi_{0,I}(x,t)\\ \psi_{1,I}(x,t)\end{array}\right)
\\=&-\frac{\mathrm{i} \lambda V(x)}{2}\left(\xi_2(t) \sigma_x+\xi_1(t) \sigma_y\right)\left(\begin{array}{l}
\psi_{0,I}(x, t) \\
\psi_{1,I}(x, t)
\end{array}\right)  \\ &+\mathrm i \frac{\lambda^{2} }{\varepsilon}
 \int_{0}^{t} \mathrm d \tau 
 \left(
 \begin{array}{c}
 c^-(\tau) V(x) e^{-\frac{\mathrm{i}}{\varepsilon}\hat h_1\tau} V(x)  \psi_{0,I}(x,t)      \\
  c^+(\tau) V(x)\mathrm e^{-\frac{\mathrm{i}}{\varepsilon}\hat h_0\tau}  V(x)\psi_{1,I}(x,t)
 \end{array}
 \right)
\end{aligned}
\label{eq:nMSSEmetal_wide_finitehistory}
\end{equation}
If we let $t_0\to-\infty$ in \cref{eq:QME_markov},  we obtain the infinite-history QME, also known as the Redfield equation:
\begin{equation}
    \begin{aligned}
   \frac{\mathrm{d} \rho_I(x,x^{\prime},t)}{\mathrm{d} t} 
    & = \frac{\lambda^2}{\varepsilon^2} 
    \left(
        \int_0^{\infty}\mathbf c_h(x,\tau)\mathrm{d}\tau\rho_I(x,x^{\prime},t)\right.\\
 + 
 \int_0^{\infty} & \left.\mathbf c(x,x^{\prime},\tau)\mathrm{d}\tau\rho_I^{(d)}(x,x^{\prime},t)
 \right)+ h.c.,\\
 \rho_I(-\infty)&=\rho_0.
\end{aligned}
\label{eq:Redfield}
\end{equation}
If we replace $\int_0^t$ with $\int_{-\infty}^0$ in \cref{eq:nMSSEmetal_wide_finitehistory}, we obtain the infinite-history SSE.
The correspondence between  SSE and QME, for example non-Markovian SSE (\cref{eq:nMSSEmetal_wide}) and non-Markovian QME (\cref{eq:QME}), finite-history SSE (\cref{eq:nMSSEmetal_wide_finitehistory})  and finite-history QME (\cref{eq:QME_markov}), infinite-history SSE and Redfield equations (\cref{eq:Redfield}) could be seen as the quantum analog of the relation between a stochastic process and its corresponding Fokker-Planck equation.

\begin{figure*}
    \centering
    \includegraphics[width=160mm]{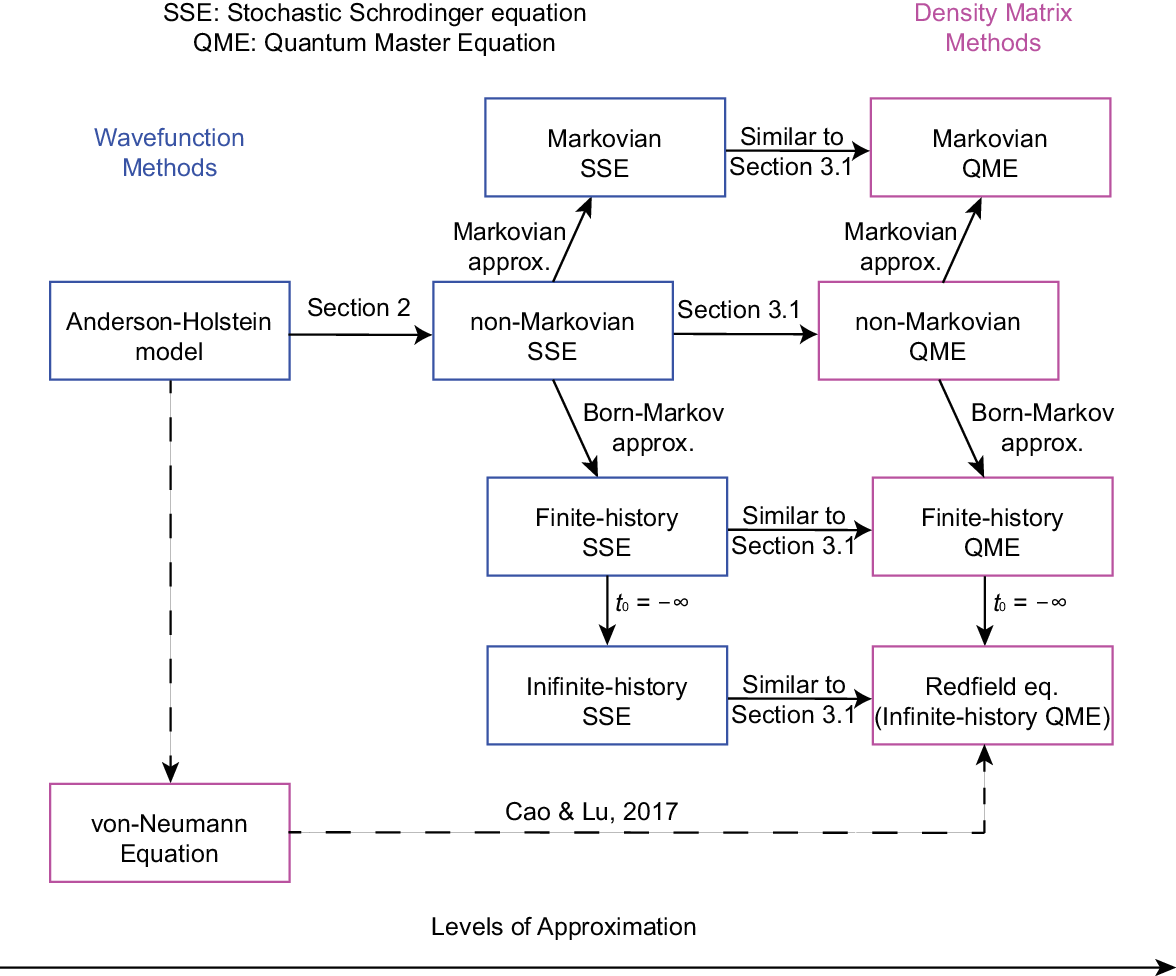}
    \caption{Hierarchy of various models for Anderson-Holstein impurities.}
    \label{fig:nMSSE}
\end{figure*}

Since QME is obtained by neglecting higher order terms in the von-Neumann type equation of SSE, it should be viewed as a further approximation on top of SSE. In other words, we have established a hierarchy of models for Anderson-Holstein impuritites, as detailed in \cref{fig:nMSSE}.
Going from SSE, to QME, and further to classical master equations (CME), one will make more  assumptions and conduct more approximations.
As a result, one would possibly lose crucial physical features while going along this hierarchy of approximations. 

\section{Numerical Methods}
\label{sec:numericalmethods}

\subsection{Noise Generation}
\label{subsec:noise}
To numerically simulate SSE, we need to generate stochastic noises that are subject to \cref{eq:correlationcontinuous}. The conventional noise generation scheme \cite{gaspard1999non} relies on the analytic continuation of the correlation function to the complex plane. Such methods suffer from  non-causality and non-physical artifacts due to the numerical instability of many rational approximation schemes \cite{trefethen2019approximation,huang2023robust}.
Here we present a stable noise generation scheme which do not rely on the interpolation of any correlation function.

We first rewrite the noise term as follows:
\begin{equation}
\begin{aligned}
-&\frac{\mathrm{i} \lambda V(x)}{2}\left(\xi_2(t) \sigma_x+\xi_1(t) \sigma_y\right)\left(\begin{array}{l}
\psi_0(x, t) \\
\psi_1(x, t)
\end{array}\right) \\ =& \lambda V(x)
\left(
\begin{array}{l}
\widetilde{\xi}_+(t)\psi_1(x, t) \\
\widetilde{\xi}_-(t)\psi_0(x, t)
\end{array}
\right),
\\ 
 \widetilde{\xi}_+(t)=&\frac{-\xi_1(t)-\mathrm{i}\xi_2(t)}{2},\quad 
        \widetilde{\xi}_-(t)= \frac{\xi_1(t)-\mathrm{i}\xi_2(t)}{2}.
    \end{aligned}
\end{equation}
where
$\tilde\xi_{\pm}(t)$ satisfies that,
\begin{equation}
    \begin{aligned}
        \mathbb{E}(\widetilde{\xi}_{\pm}(t))=0,&\quad
  \mathbb{E}(\widetilde{\xi}_+^{*}(t)\widetilde{\xi}_-(t'))=0,\\\mathbb{E}(\widetilde{\xi}_{\pm}(t)\widetilde{\xi}_{\pm}(t'))&=\mathbb{E}(\widetilde{\xi}_{\pm}(t)\widetilde{\xi}_{\mp}(t'))=0, \quad  \\
        \mathbb{E}(\widetilde{\xi}_+^{*}(t)\widetilde{\xi}_{+}(t'))&=c^{+}(t-t'),\\
        \mathbb{E}(\widetilde{\xi}_-^{*}(t)\widetilde{\xi}_{-}(t'))&=c^{-}(t-t').
    \end{aligned}
\end{equation}
In this way, $\tilde \xi_+(t)$ and $\tilde\xi_-(t)$ are decoupled, and could be generated separately.
Let us define
$$
W_{\pm}(t) = \int_0^t \tilde \xi_{\pm}(\tau)\mathrm d\tau,
$$
Then $W_{\pm}(t)$ is a Gaussian process with the covariance function $K(t,s)$:
$$
K_{\pm}(s,t) = \int^{s}_0\mathrm{d}\tau_1\int^{t}_0\mathrm{d}\tau_2 c^{\pm}(\tau_1-\tau_2).
$$
In particular, when $c^{\pm}(\tau_1-\tau_2) = \delta(\tau_1-\tau_2)$, we have $K_{\pm}(s,t) = \min(s,t)$, and $W_{\pm}(t)$ is the standard Brownian motion.

Generally, the sampling of $W_{\pm}(t)$ can be achieved by the  Karhunen-Lo\`{e}ve expansion. For a specified maximum time $T_{\text{max}}$, consider the following eigenvalue problem:
\begin{equation}
\int_0^{T_{\text{max}}} K_{\pm}(s, t) \phi^{\pm}_i(t) \mathrm d t=\lambda^{\pm}_i\phi^{\pm}_i(s), \quad i=1,2, \ldots
\label{eq:eigenvalue_kernel}
\end{equation}
with the normalization condition $\int_0^{T_{\text{max}}} \phi_i^{\pm} \phi_j^{\pm} \mathrm d t=\delta_{i j}$. Then $W_{\pm}(t)$ has the Karhunen-Lo\`{e}ve expansion
\begin{equation}
   W_{\pm}(t)=\sum_{k=1}^{\infty} \alpha_k^{\pm}\sqrt{\lambda_k^{\pm}} \phi_k^{\pm}(t)
\end{equation}
Here the $\alpha_k \sim N(0,1)$ are i.i.d. random variables. 
In this way, the task of sampling  the time-dependent noise is reduced to sampling scalar time-independent random variables. The eigenvalue problem \cref{eq:eigenvalue_kernel} could be solved with a finite difference discretization in the time variable.
\subsection{Time evolution}
\label{subsec:timeevol} 
We are solving  \cref{eq:nMSSEmetal_wide} on the spatial domain $[a,b]$ and for time $[0,T]$. We use the following discretization:
\begin{equation}
    \begin{aligned}
              \Delta x=\dfrac{(b-a)}{M},\quad   &
      \Delta t=\dfrac{T}{N}.\\
      x_j:=a+j\Delta x,\quad &j=0,1,\cdots,M-1,\\   t_n:=n\Delta t,\quad &n=0,1,2\cdots,N. 
    \end{aligned}
\end{equation}
To  numerically solve \cref{eq:nMSSEmetal_wide}, our strategy is as follows:
\begin{itemize}
    \item For the system Hamiltonian $\hat H_S$, we use the Time Splitting Spectral method \cite{bao2002time};
    \item For the fluctuation and dissipation terms, we use the Euler-Maruyama algorithm \cite{kloeden1992stochastic}.
\end{itemize}
Let us present the numerical algorithm in detail for the Markovian SSE. With an efficient evaluation of the integration term,  the strategy we described below   is applicable to the non-Markovian case.
In the Markovian limit, let $c_{\pm}(t) $ becomes $c_0^{\pm}\delta(t)$, where $c_0^{\pm}$ are fixed constant. Then \cref{eq:nMSSEmetal_wide} becomes
\begin{equation}
\begin{aligned}
&\mathrm{i}\varepsilon
\left(
\begin{array}{c}
\mathrm{d} \psi_0(x,t)\\ 
\mathrm{d} \psi_1(x,t)
\end{array}
\right)\\
=&
\left(
\begin{array}{c}
\hat{h}_0\psi_0(x,t)\\ 
\hat{h}_1\psi_1(x,t)
\end{array}
\right)\mathrm{d} t
+\lambda V(x)
\left(
\begin{array}{l}
\psi_1(x, t)\mathrm{d}W_1(t)\\
\psi_0(x, t)\mathrm{d}W_0(t)
\end{array}
\right) \\
&+\frac{\mathrm i \lambda^{2}}{\varepsilon}|V(x)|^2
\left(\begin{array}{c}
   c_0^-  \psi_0(x,t)  \\
     c_0^+\psi_1(x,t)
\end{array}\right)\mathrm{d} t,\quad x\in[a,b],
\end{aligned}
\label{eq:MSSEmetal}
\end{equation}
Let us consider the following initial condition, which means that at $t=0$ the molecule is neutral:
$$
\left(
\begin{array}{c}
\psi_0(x,0)\\ 
\psi_1(x,0)
\end{array}
\right)
=
\left(
\begin{array}{c}
\psi_0(x,0)\\ 
0
\end{array}
\right).$$

 Denote the numerical solution of $\psi_{k}(x_j,t_n),k=0,1$ by $\psi_{k}^{j,n},k=0,1$. From time $t=t_n$ to time $t=t_{n+1}$, we do the following:
\begin{enumerate}
    \item Evolve using the potential term for a time step:
     \begin{equation}
    \begin{aligned}
            \psi_{k}^{j,*1} 
        =
            \exp{\left(-\mathrm{i}\frac{U_k(x_j)}{\varepsilon}\Delta t\right)}\psi_{k}^{j,n} ,\\ k=0,1,\quad 
    j=0,\cdots,M-1,  
    \end{aligned}
    \end{equation}
\item Evolve using the kinetic term for a time step, but in the frequency domain: for $k=0,1$, we have
\begin{equation}
    \begin{aligned}
        \begin{cases}
        \{\widehat{\psi}_{k}^{l,*1}\}_{l=-\frac{M}{2}}^{\frac{M}{2}-1} = \mbox{FFT}\left(\{\psi_{k}^{j,*1}\}_{j=0}^{M-1}\right)
    ,
      \\
      \widehat{\psi}_{k}^{l,*2} =  \widehat{\psi}_{k}^{l,*1}\exp{\left(-\mathrm{i}\varepsilon\frac{\mu_l^2}{2}\Delta t\right)},\text{ } 
            l=-\frac{M}{2},\cdots,\frac{M}{2}-1,\\
\{\psi_{k}^{j,*2}\}_{j=0}^{M-1} = \mbox{iFFT}
   \left(   \{\widehat{\psi}_{k}^{l,*2}\}_{l=-\frac{M}{2}}^{\frac{M}{2}-1}   \right) .   
       \end{cases}
    \end{aligned}
\end{equation}
Here $\mu_l=\dfrac{2\pi l}{b-a}$, FFT and iFFT denote the Fast Fourier Transform and its inverse.
\item Deal with the fluctuation-dissipation term using the Euler-Maruyama scheme:
\begin{equation}
    \begin{aligned}
     &   \left(\begin{array}{l}
            \psi_{0}^{j,n+1} \\
            \psi_{1}^{j,n+1}
        \end{array}\right)\\
        &=
        \left(\begin{array}{l}
            \psi_{0}^{j,*2} \\
            \psi_{1}^{j,*2}
        \end{array}\right)
        -\mathrm{i}\dfrac{\lambda V(x_j)}{\varepsilon}
        \left(\begin{array}{l}
            \left(W_1(t_{n+1})-W_1(t_n)\right)\psi_{1}^{j,*2} \\
            \left(W_0(t_{n+1})-W_0(t_n)\right)\psi_{0}^{j,*2}
        \end{array}\right)\\
        &+\dfrac{\lambda^2}{\varepsilon^2}|V(x_j)|^2
        \left(\begin{array}{l}
            c_0^-\psi_{0}^{j,*2} \\
            c_0^+\psi_{1}^{j,*2}
        \end{array}\right)\Delta t.
    \end{aligned}
\end{equation}
\end{enumerate}

\begin{remark}
    Here we use a first-order splitting scheme for the Hamiltonian part, which is sufficient for our problem. Higher-order splitting schemes also exist, such as Strang-Splitting and so on.
\end{remark}

\section{Numerical Results}
\label{sec:numericalresults}
The purpose of our numerical experiments is two-fold. On the one hand,  we demonstrate that an ensemble of realizations of SSE could be used to obtain samples of physical observables of interest and thus directly manifest distributional information of these quantities, while QME could not provide a detailed characterization of the distribution except for  moment information. On the other hand, we show that SSE and QME approaches indeed have the same thermodynamic equilibrium, while exhibiting different transient dynamics towards reaching such equilibrium.

\subsection{Samples of physical observables by SSE}
Let $x\in [a,b] = [-\pi,\pi]$, the maximal time $T=10$. For simplicity, in this subsection, we choose the interaction strength $\lambda=\varepsilon$, which is neither necessary nor essential. 
Let us focus on the following system potentials which are harmonic with different centers: 
$$
U_0(x) = \frac{1}{2}x^2, \quad U_1(x) = \frac{1}{2}x^2+ 0.1x.
$$
We aim to numerically explore SSE \eqref{eq:MSSEmetal} in the wide band limit and with $\delta$-correlated noise with special attention to investigating the role of the coupling potential $V(x)$. To this end, we consider the following examples:

\textbf{Example 1}. Propagating a Gaussian wavepacket with a bimodal coupling function:
\begin{equation}
\begin{aligned}    \psi_0(x,0)&=\dfrac{1}{(\pi\varepsilon)^{\frac{1}{4}}}\exp{\left(\frac{-(x-q_0)^2}{2\varepsilon}+\mathrm{i}\frac{p_0(x-q_0)}{\varepsilon}\right)},\\
V(x)&=\exp{\left(-10(x-0.5)^2\right)}+\exp{\left(-40(x+2)^2-1\right)},\\
q_0&=-1,p_0=0.5,
\end{aligned}
\end{equation}

\textbf{Example 2}. Propagating a Gaussian wavepacket with another bimodal coupling function:
\begin{equation}
\begin{aligned}
    \psi_0(x)&=\dfrac{1}{(\pi\varepsilon)^{\frac{1}{4}}}\exp{\left(\frac{-(x-q_0)^2}{2\varepsilon}+\mathrm{i}\frac{p_0(x-q_0)}{\varepsilon}\right)},\\
    V(x)&=2\exp{\left(-10(x-0.9)^2\right)}+5\exp{\left(-40(x+0.5)^2\right)},\\
    q_0&=-1,p_0=0.5.
\end{aligned}
\end{equation}

\begin{figure*}[htbp]
    \centering
\subfigure{
\includegraphics[width=0.3\linewidth]{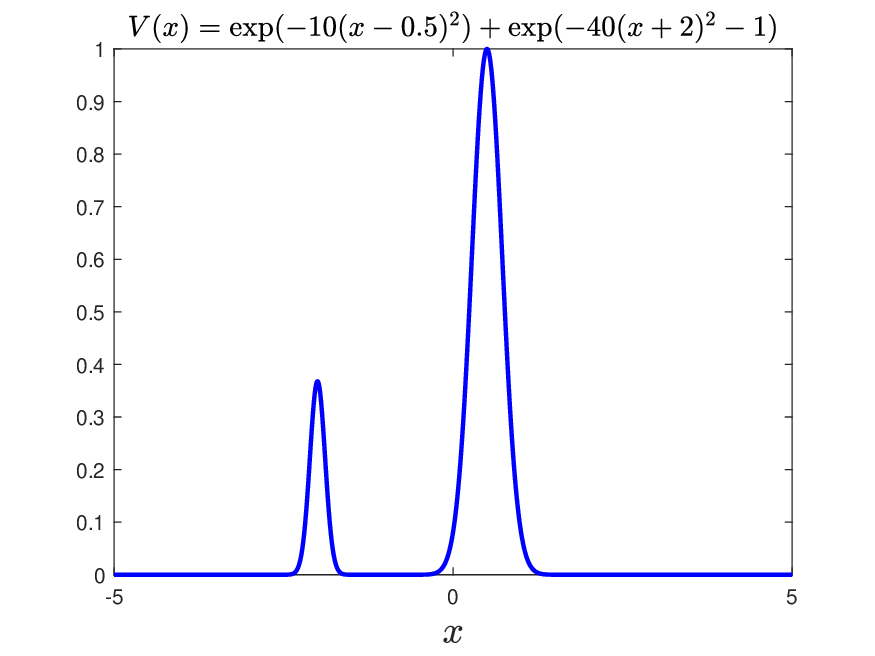}}
\subfigure{
\includegraphics[width=0.3\textwidth]{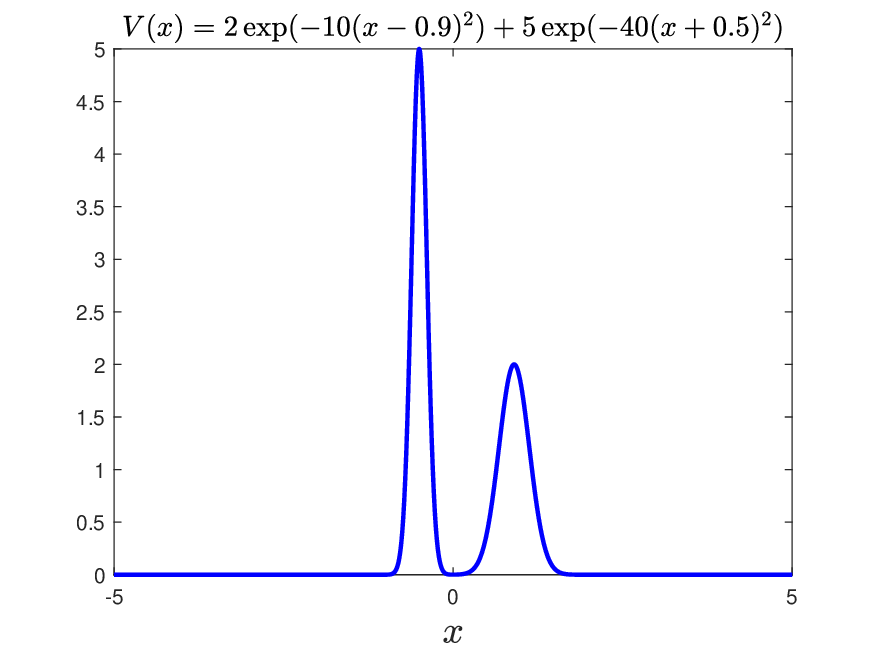}
}
\subfigure{
\includegraphics[width=0.3\textwidth]{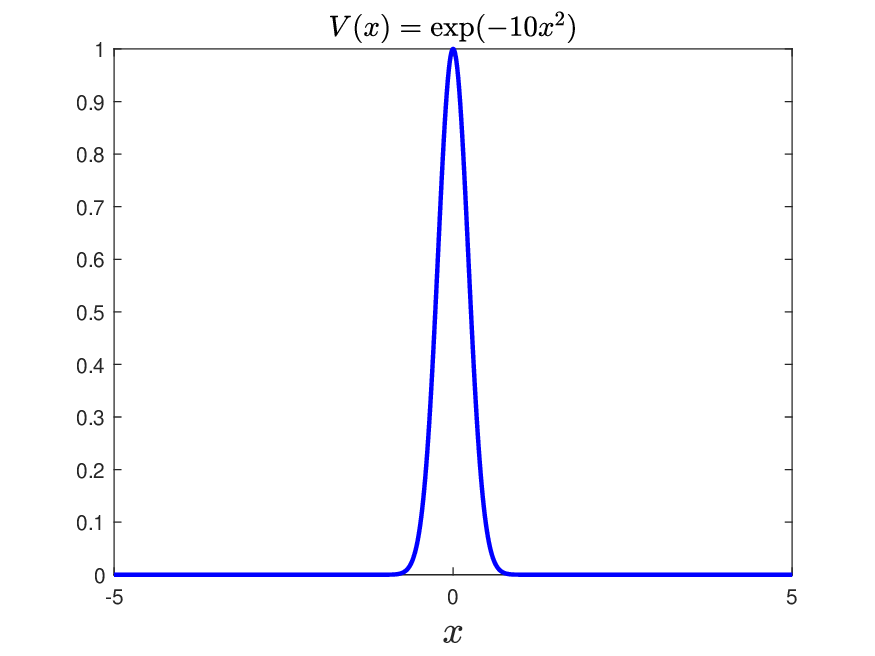}}
    \caption{Plots of coupling functions $V(x)$ in example 1, 2 and 3.}\label{fig:Vx}
\end{figure*}

\textbf{Example 3}. Propagating a non-Gaussian wavepacket with a unimodal coupling function:
\begin{equation}
\begin{aligned}    \psi_0(x)&\propto\exp{\left(-5(x+1)^2+\mathrm{i}\frac{\sin(x)}{\varepsilon}\right)},\\
V(x)&=\exp{\left(-10x^2\right)}.
\end{aligned}
\end{equation}
The coupling functions $V(x)$ that we considered are shown in \cref{fig:Vx}. We remark that the experiments are designed such that either the wave packet is expected to exhibit decoherence through the interaction with the bath or is not initialized from a coherent state. 

\begin{figure*}[htbp]
    \centering
    \subfigbottomskip=2pt
    \subfigcapskip=-5pt
    \subfigure{
		\includegraphics[width=0.32\linewidth]{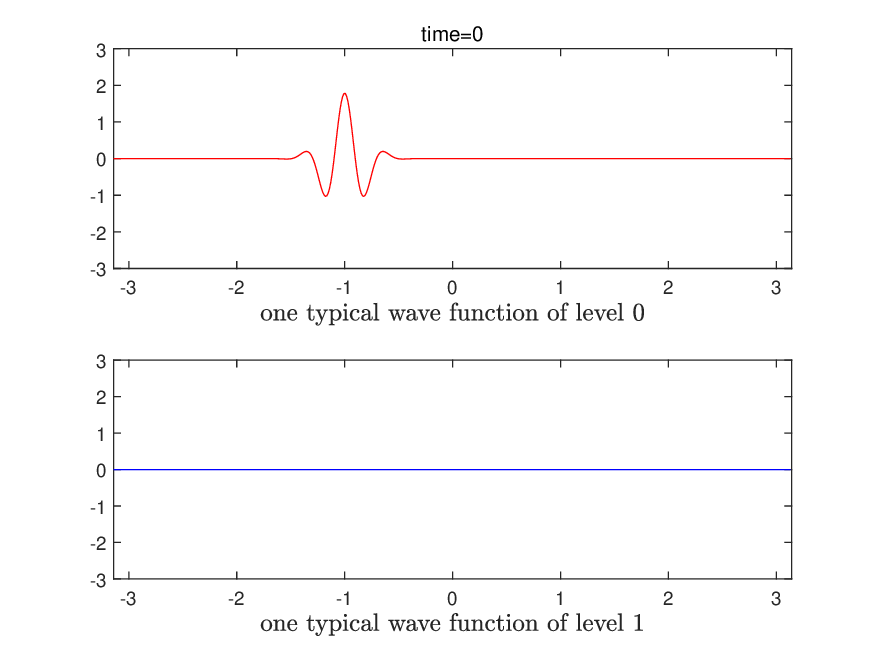}}
    \subfigure{
		\includegraphics[width=0.32\linewidth]{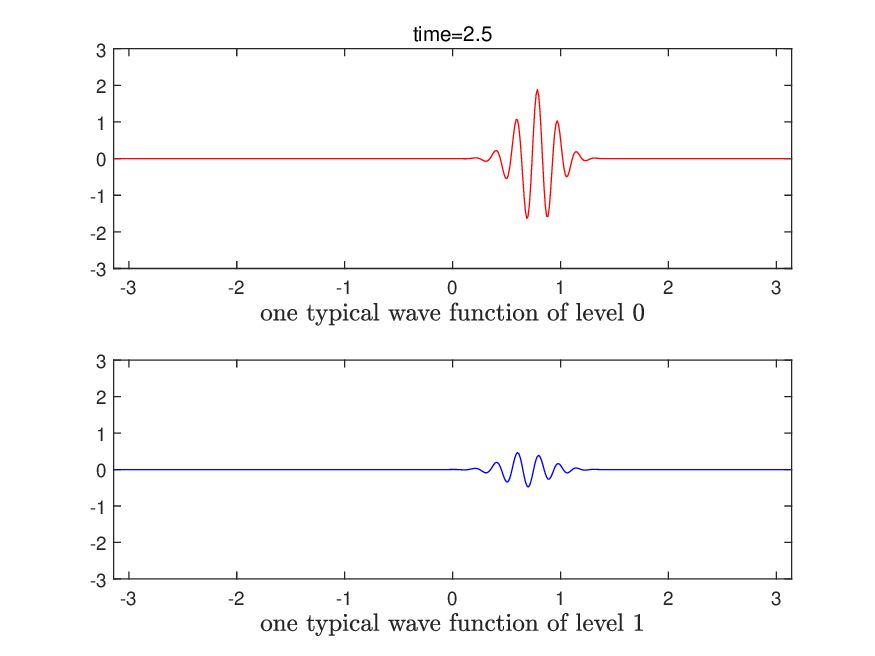}}
    \subfigure{
		\includegraphics[width=0.32\linewidth]{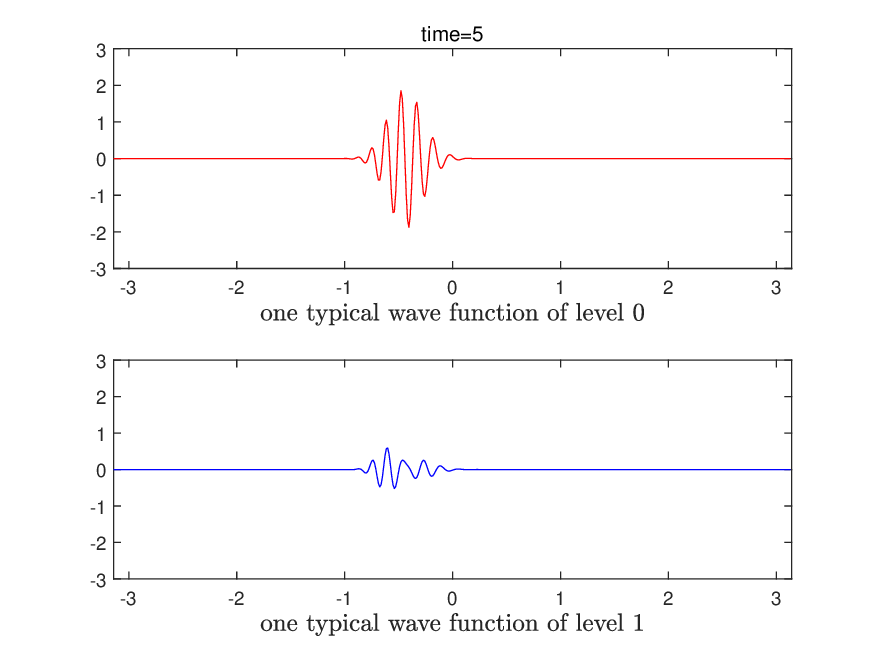}}
    \\
    \subfigure{
		\includegraphics[width=0.32\linewidth]{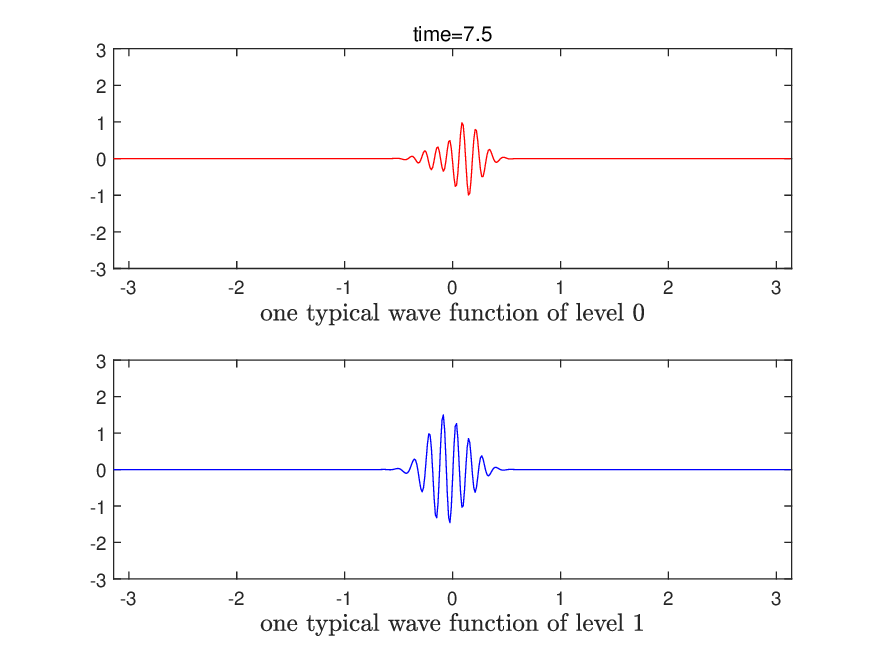}}
    \subfigure{
		\includegraphics[width=0.32\linewidth]{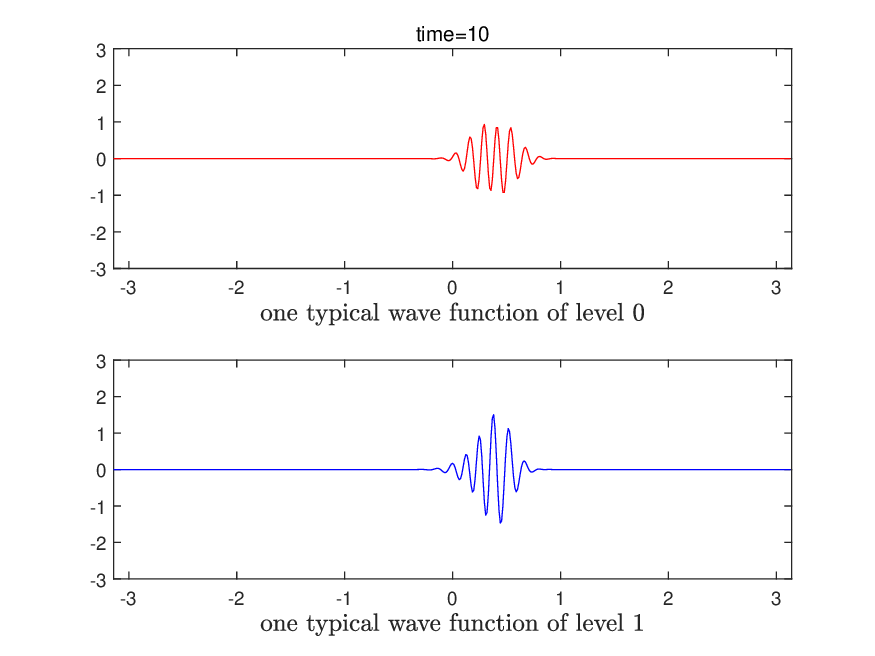}}
    \caption{The evolution of a typical wavefunction trajectory in example $1$.}
    \label{fig: typicaltrajectory}
\end{figure*}

A typical trajectory of time evolution of SSE is shown in \cref{fig: typicaltrajectory}. We observe that the wave function is initially populated solely on level-0, it oscillates and propagates within a finite region due to the confinement of the harmonic potentials. In particular, only when it passes through an interaction region (where the coupling potential $V(x)$ is significant), it will partially and stochastically transmit to the other level. Thus the resulting dynamical behavior is rather complicated and the nonadiabatic phenomenon is nontrivial.

\begin{figure*}[htbp]
	\centering  
	\subfigbottomskip=2pt 
	\subfigcapskip=-5pt 
	\subfigure{
\includegraphics[width=0.24\linewidth]{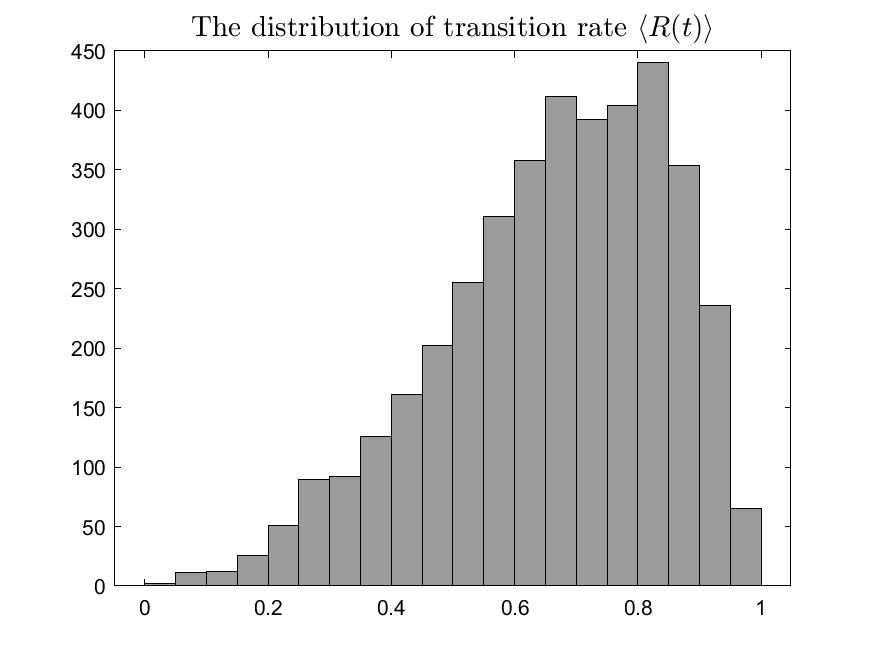}}
\subfigure{	\includegraphics[width=0.23\linewidth]{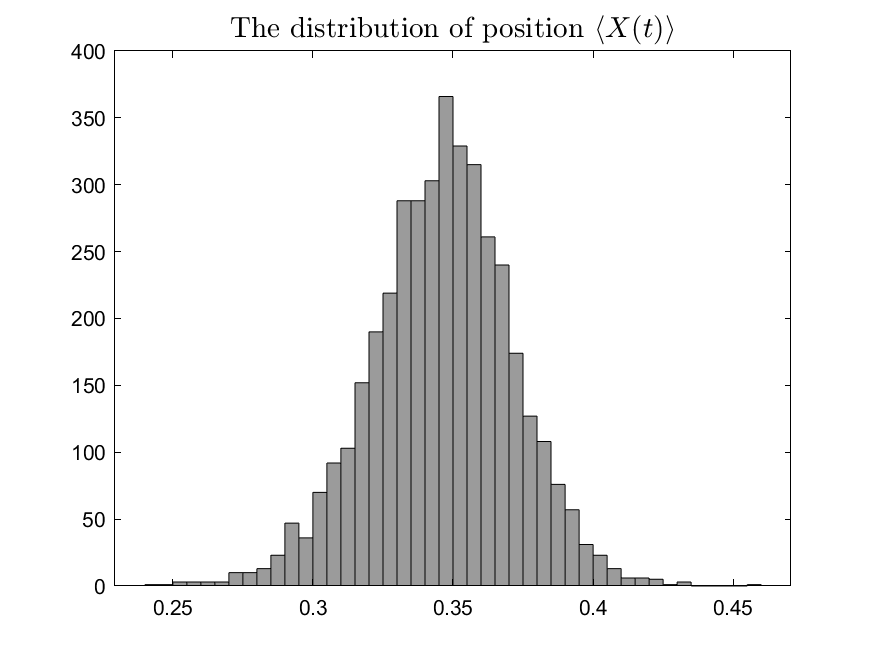}}
\subfigure{
\includegraphics[width=0.24\linewidth]{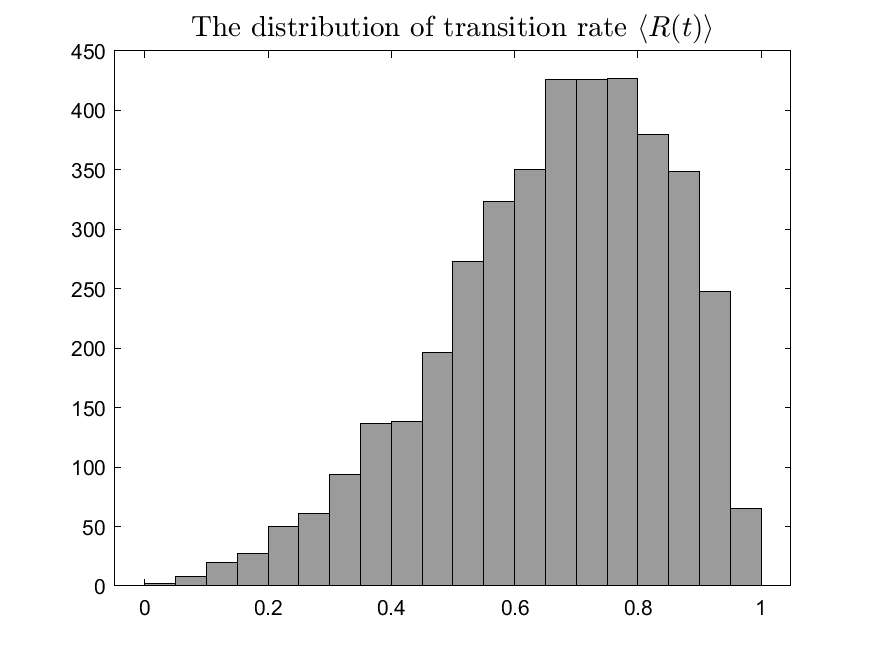}}
\subfigure{
\includegraphics[width=0.23\linewidth]{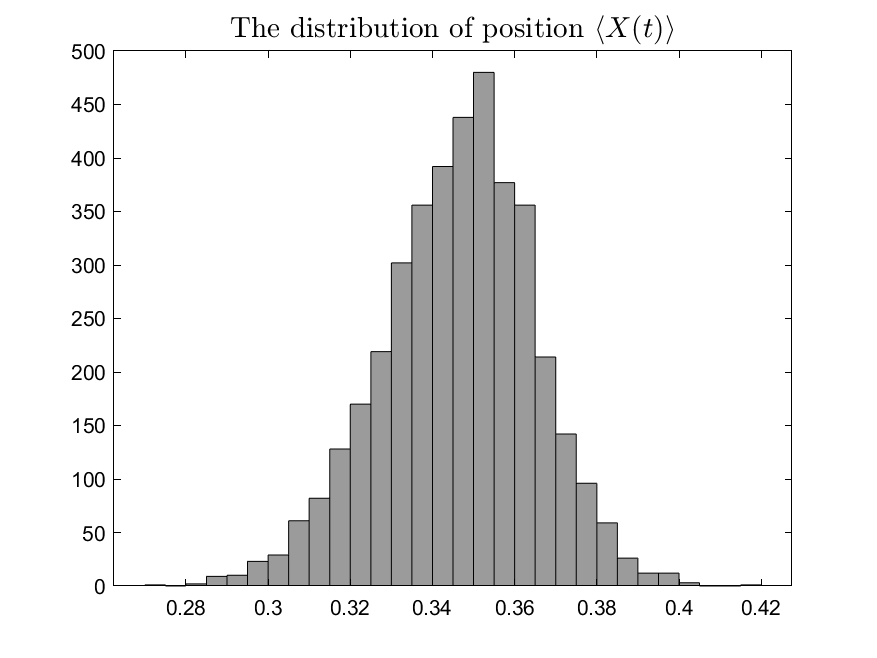}}
	\caption{Histogram 
 for transition rate and atomic position at $T=10$, with $\varepsilon=1/32$ and $\varepsilon= 1/64$, in example 1, from 4000 trajectories. The two subfigures on the left correspond to $\lambda=1/32$ and the ones on the right correspond to $\lambda=1/64$. This demonstrates a  transition rate with a non-Gaussian distribution but an atomic position with a Gaussian-like distribution. }
    \label{fig: transition rate and position of Example 1}
\end{figure*}

\begin{figure*}[htbp]
	\centering  
	\subfigbottomskip=2pt 
	\subfigcapskip=-5pt 
	\subfigure{
	\includegraphics[width=0.23\linewidth]{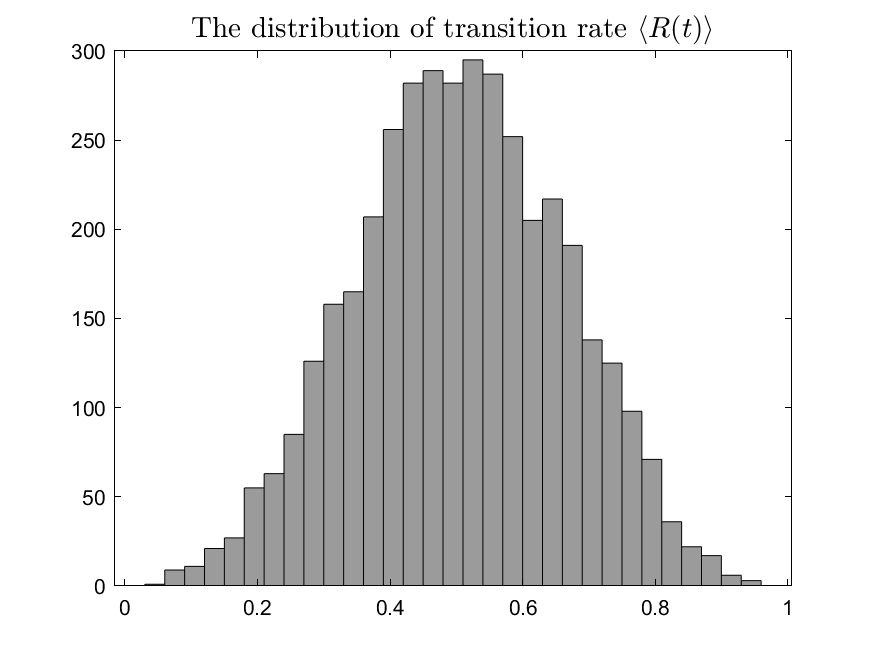}}
	\subfigure{
		\includegraphics[width=0.23\linewidth]{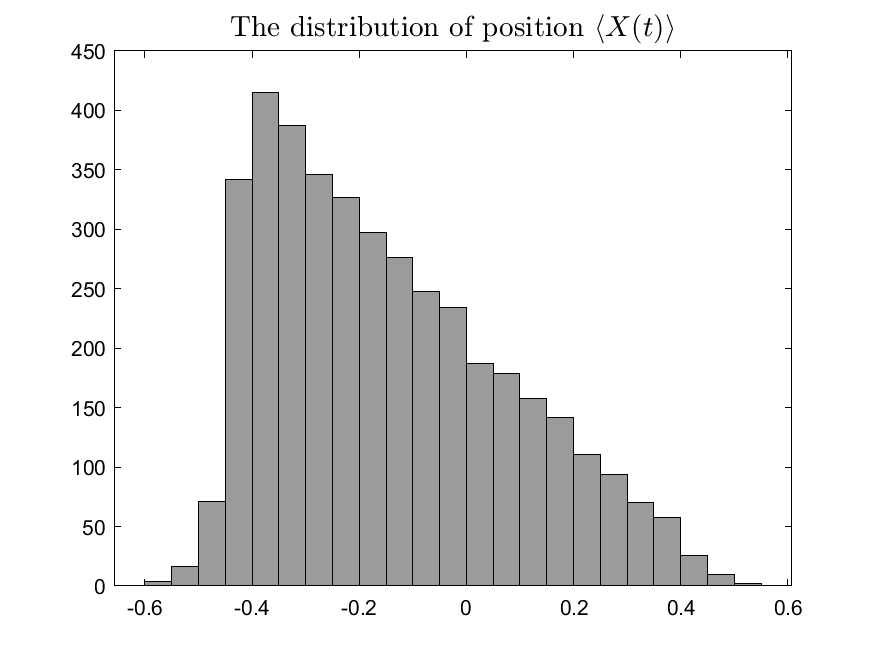}}
	\caption{Histogram 
 for transition rate and atomic position at $T=10$, with $\varepsilon=1/32$, in example 2, from 4000 trajectories. This demonstrates a  transition rate with a Gaussian-like distribution but an atomic position with a non-Gaussian distribution.}
    \label{fig: transition rate and position of Example 2}
\end{figure*}

\begin{figure*}[htbp]
	\centering  
	\subfigbottomskip=2pt 
	\subfigcapskip=-5pt 
	\subfigure{
\includegraphics[width=0.24\linewidth]{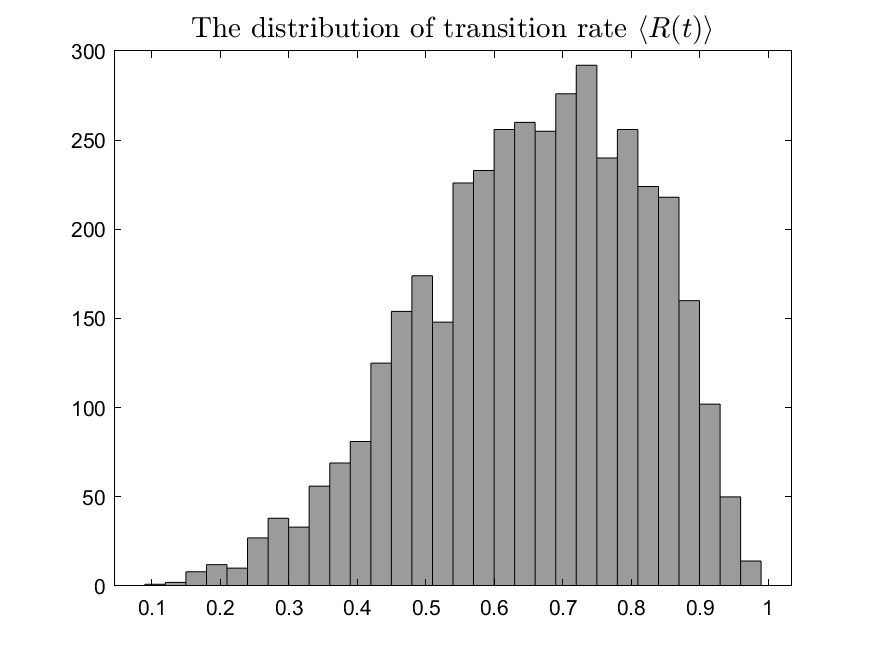}}
\subfigure{	\includegraphics[width=0.23\linewidth]{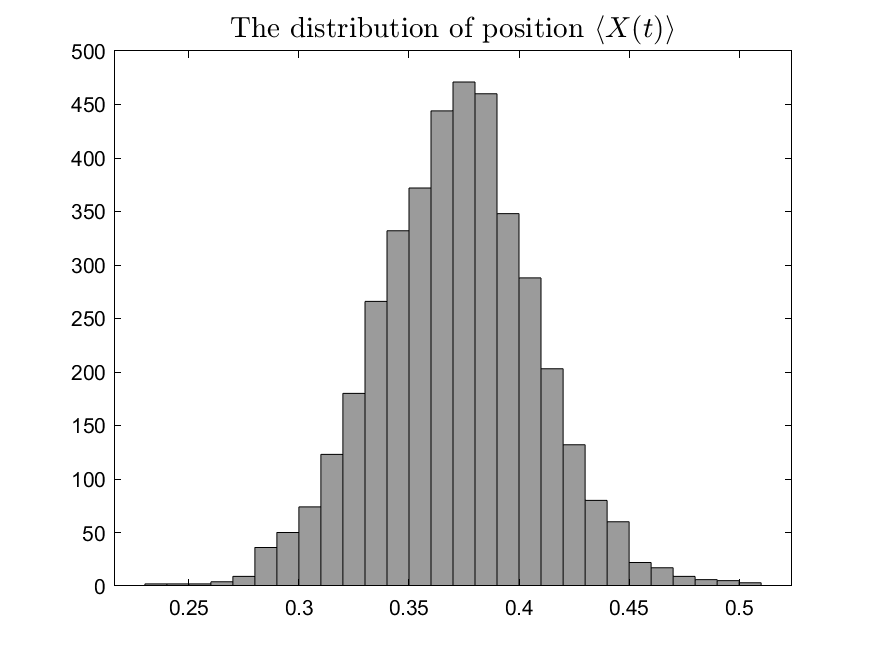}}
\subfigure{
\includegraphics[width=0.24\linewidth]{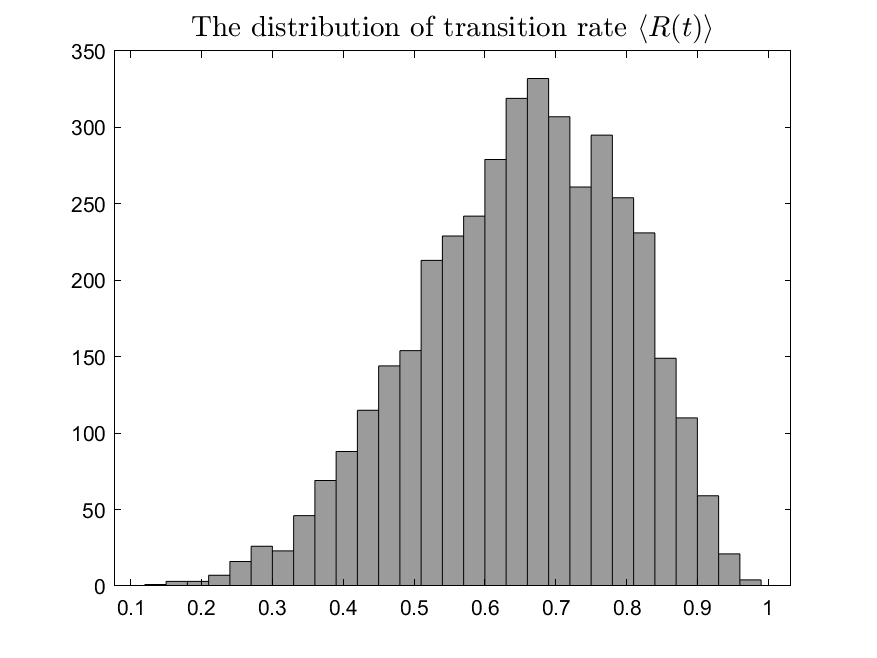}}
\subfigure{
\includegraphics[width=0.23\linewidth]{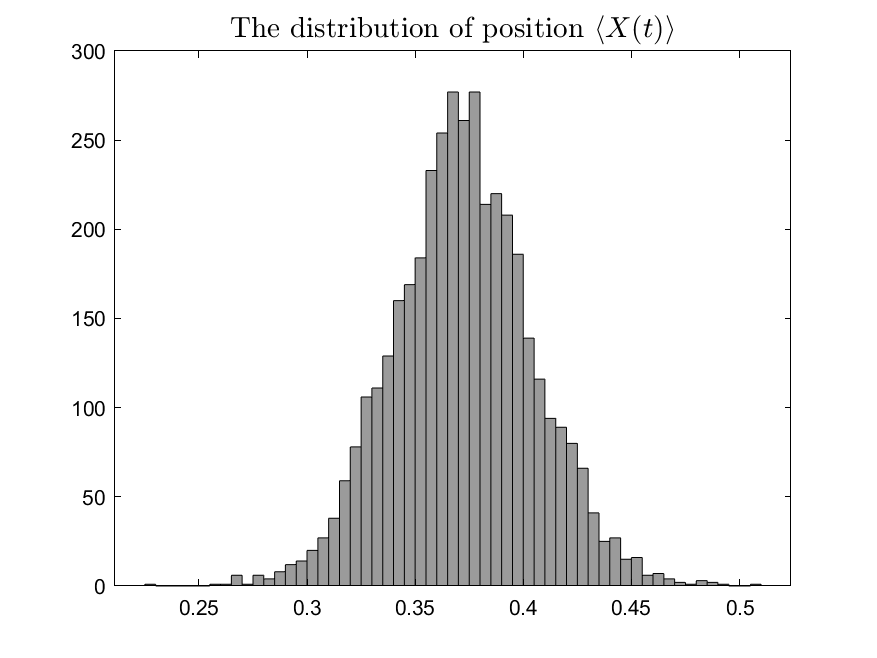}}
\caption{Histogram 
 for transition rate and atomic position at $T=10$, with $\varepsilon=1/32$ and $\varepsilon= 1/64$, in example 3, from 4000 trajectories. The two subfigures on the left correspond to $\lambda=1/32$ and the ones on the right correspond to $\lambda=1/64$. This shows that the transition rate and atomic position  both have a Gaussian-like distribution.}
    \label{fig: transition rate and position of Example 3}
\end{figure*}

To explore the stochasticity of the SSE model, let us demonstrate with the numerical simulations of certain physical observables computed via an ensemble of realizations of SSE \eqref{eq:MSSEmetal}. We consider the transition rate $\langle R(t)\rangle$ and spatial average $\langle X(t)\rangle$, defined as: 
\begin{equation}
    \begin{aligned}
        \langle R(t)\rangle=\frac{\int|\psi_{0}(x,t)|^2\mathrm{d}x}{\int\left(|\psi_{0}(x,t)|^2+|\psi_{1}(x,t)|^2\right)\mathrm{d}x},\\
        \langle X(t)\rangle=\frac{\int x\left(|\psi_{0}(x,t)|^2+|\psi_{1}(x,t)|^2\right)\mathrm{d}x}{\int\left(|\psi_{0}(x,t)|^2+|\psi_{1}(x,t)|^2\right)\mathrm{d}x},
    \end{aligned}
\end{equation}
We emphasize that $ \langle \cdot \rangle$ mean taking the average over the quantum state, which is obtained by a one-time realization of SSE \eqref{eq:MSSEmetal}. Hence, $\langle R(t)\rangle$ and $\langle X(t)\rangle$ are random variables due to the stochasticity in the dynamics of \eqref{eq:MSSEmetal}. We repeat simulation \eqref{eq:MSSEmetal} of each exmaple for $4000$ times, and the statistics (in the form of the histogram) of the observables are shown in    \cref{fig: transition rate and position of Example 1}, \cref{fig: transition rate and position of Example 2} and \cref{fig: transition rate and position of Example 3}, respectively.

 We can see that the coupling function $V(x)$ significantly affects the probability distribution of physical observables and SSE is capable of capturing both Gaussian and non-Gaussian distributional information of these quantities.
Specifically, we have the following observations:
\begin{itemize}
    \item[(1)] In Example 1, the coupling function $V(x)$ is bimodal: the primary coupling peak is placed near the center of the propagation region and the secondary coupling peak is placed near the boundary of the propagation region, where wave function turns its moving direction. We observe that the distribution of transition rate $\langle R(t)\rangle$ is non-Gaussian but the distribution of atomic position $\langle X(t)\rangle$ is Gaussian;
    \item[(2)] In Example 2, the coupling function $V(x)$ is also bimodal: both coupling peaks are near the center of the propagation region, but they are of different profiles. We observe that the distribution of transition rate $\langle R(t)\rangle$ is Gaussian but the distribution of atomic position $\langle X(t)\rangle$ is non-Gaussian;
    \item[(3)] In Example 3, although the wave function is initialized far from a coherent state, the coupling function $V(x)$ is unimodal, we have the distributions of transition rate $\langle R(t)\rangle$ and atomic position $\langle X(t)\rangle$ are both Gaussian.
\end{itemize}
While the results above are not conclusively definitive, this set of experiments already sheds light on the diverse stochastic behavior contained within the SSE models.

\subsection{Comparison between SSE and QME}
This part is devoted to a  detailed comparison between SSE and QME models both in dynamics and equilibrium.
Consider \cite{cao_lindblad_2017,dou2016broadened,dou_surface_2015}
\begin{equation} 
     \begin{aligned}
        U_0(x)&=\frac{1}{2}x^2,\quad U_1(x)=U_0(x)+\sqrt{2}gx+g^2+E_d,
    \end{aligned}
\end{equation}
with the Markovian assumption $c^{\pm}(\tau) = c_0^{\pm}\delta(\tau)$,
the corresponding QME of \cref{eq:QME}, transformed back to the Schr\"{o}dinger picture, is
\begin{equation}
    \begin{aligned}
     & \frac{\mathrm{d} \rho(t)}{\mathrm{d} t} 
     =-\frac{\mathrm{i}}{\varepsilon} \left[\hat H_S, \rho(t)\right]\\+&\frac{\lambda^2}{\varepsilon^2} |V|^2
      \begin{pmatrix}
2(\rho_{00}(t)-\rho_{11}(t))&\rho_{01}(t)+\rho_{10}(t)\\
\rho_{01}(t)+\rho_{10}(t)&2(\rho_{11}(t)   -\rho_{00}(t)   )     \end{pmatrix}.
\end{aligned}
\end{equation}
This is equivalent to the traditional Markovian QME
\begin{equation}
    \begin{aligned}
&\mathrm{i}\varepsilon\frac{\mathrm{d}\rho(t)}{\mathrm{d}t}=
        \left[\begin{pmatrix}
            \hat{h}_0&0\\
            0&\hat{h}_1
        \end{pmatrix},\rho(t)\right]\\-&\mathrm{i}\frac{\lambda^2 V^2}{\varepsilon}
        \left[\hat{d}\hat{d}^{\dagger}\rho(t)-\hat{d}^{\dagger}\rho(t)\hat{d}+\rho(t)\hat{d}^{\dagger}\hat{d}-\hat{d}\rho(t)\hat{d}^{\dagger}+h.c.\right],
    \end{aligned}
\end{equation}
 where $\hat{d}$ and $\hat{d}^{\dagger}$ are defined as in \eqref{eq:ddirac}.
We choose the following parameter:
\begin{equation}
    \begin{aligned}
        L=20,\quad g=0.1,\quad 
 E_d=0.1,\quad V(x)=1,
    \end{aligned}
\end{equation}
and
the initial value
$$
\psi_0(x,0) \propto 1, \quad \rho(x,x',0) = \psi_0(x,0) \psi^*_0(x',0).
$$
In this example, we conduct the spatial discretization with Hermite polynomials. Also, note that the coupling function $V$ is chosen to be spatially homogeneous, and the system is expected to reach the equilibrium.

Recall that the proposed models in this work are characterized by two parameters: the interaction strength $\lambda$ and the semiclassical parameter $\varepsilon$. From the derivation of QME based on SSE, we can see that it holds when $\lambda \ll 1$ while $\varepsilon$ is fixed, and a recent mathematical study reveals that QME can be directly derived only if $\lambda \ll \varepsilon$ \cite{cao_lindblad_2017}. In fact, our formulation facilitates more systematic comparisons between SSE and QME with various combinations of $\lambda$ and $\varepsilon$. The purpose of the following two numerical experiments is to compare the dynamics and steady states between SSE and QME under different $\varepsilon$ versus $\lambda$ relationships. 

\begin{figure*}[htbp]
    \centering
    \subfigbottomskip=2pt 
    \subfigcapskip=-5pt 
    \subfigure{
		\includegraphics[width=0.23\linewidth]{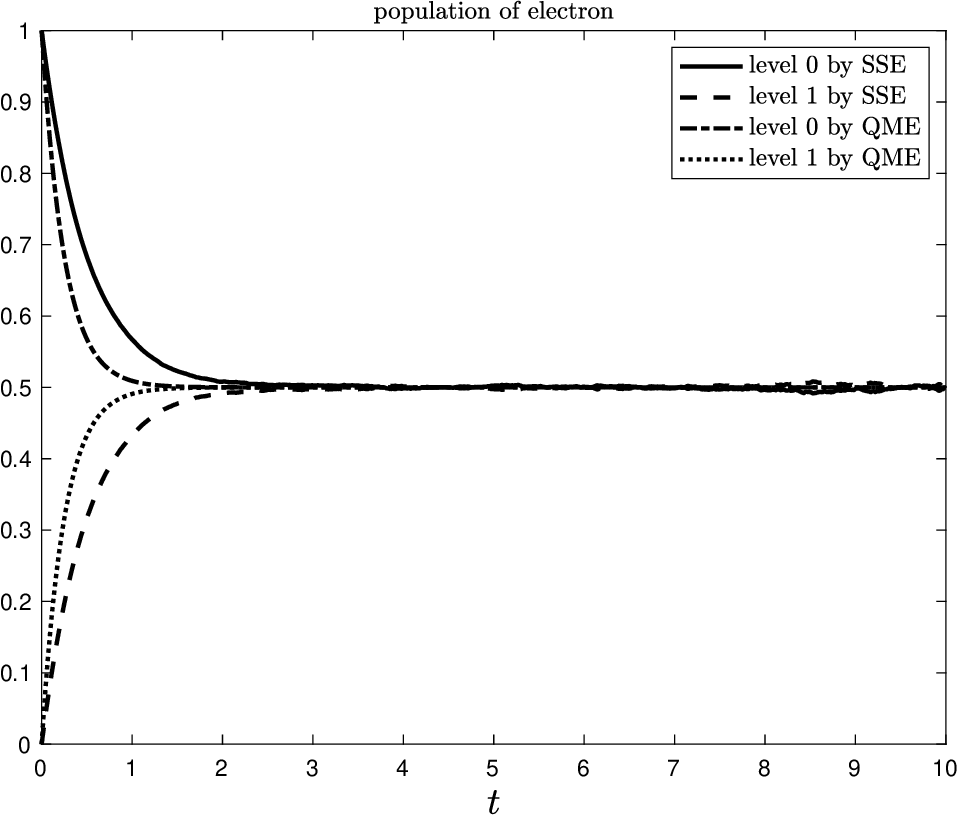}}
    \subfigure{
		\includegraphics[width=0.23\linewidth]{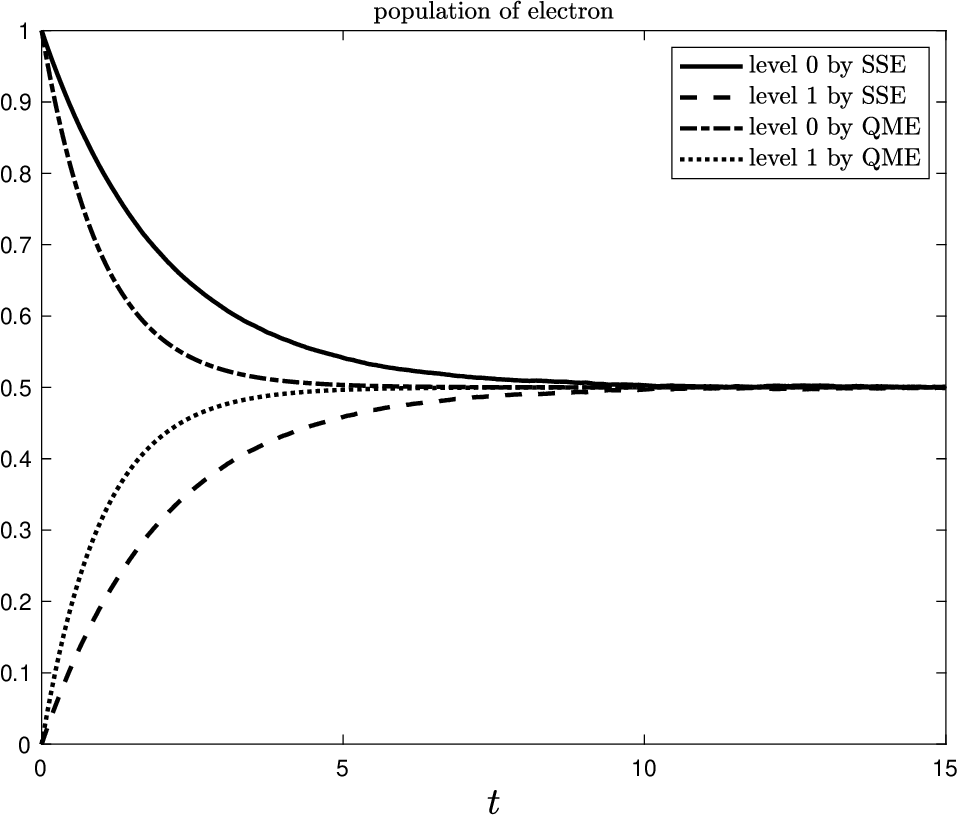}}
    \subfigure{
		\includegraphics[width=0.23\linewidth]{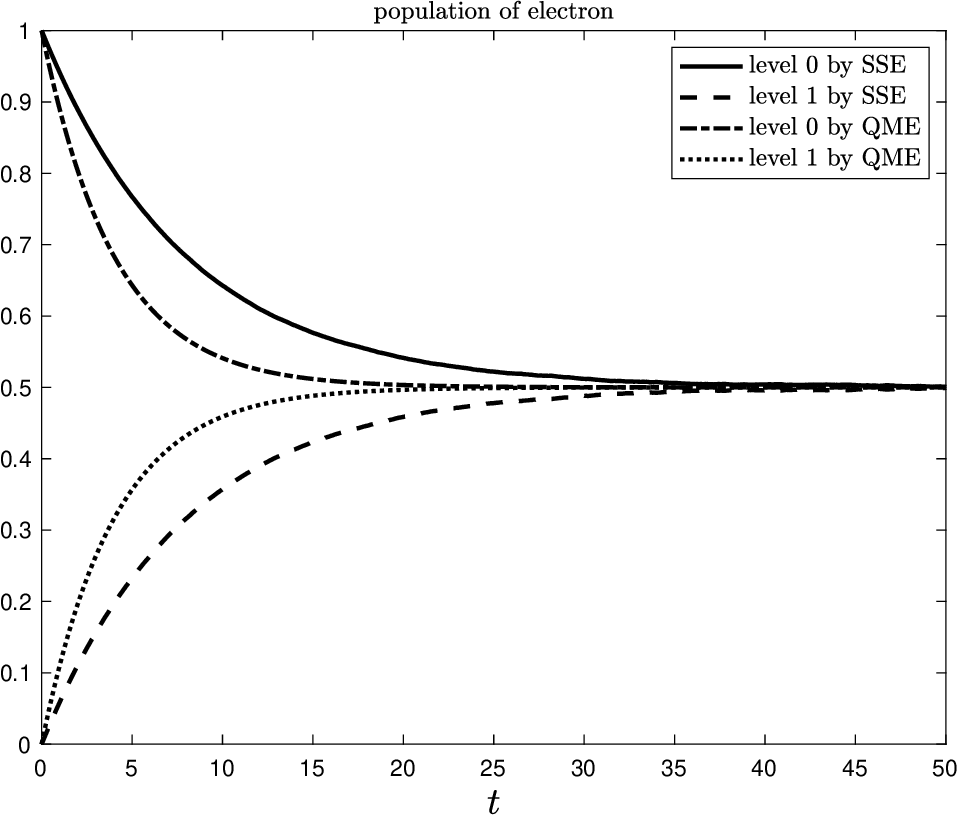}}
    \subfigure{
		\includegraphics[width=0.23\linewidth]{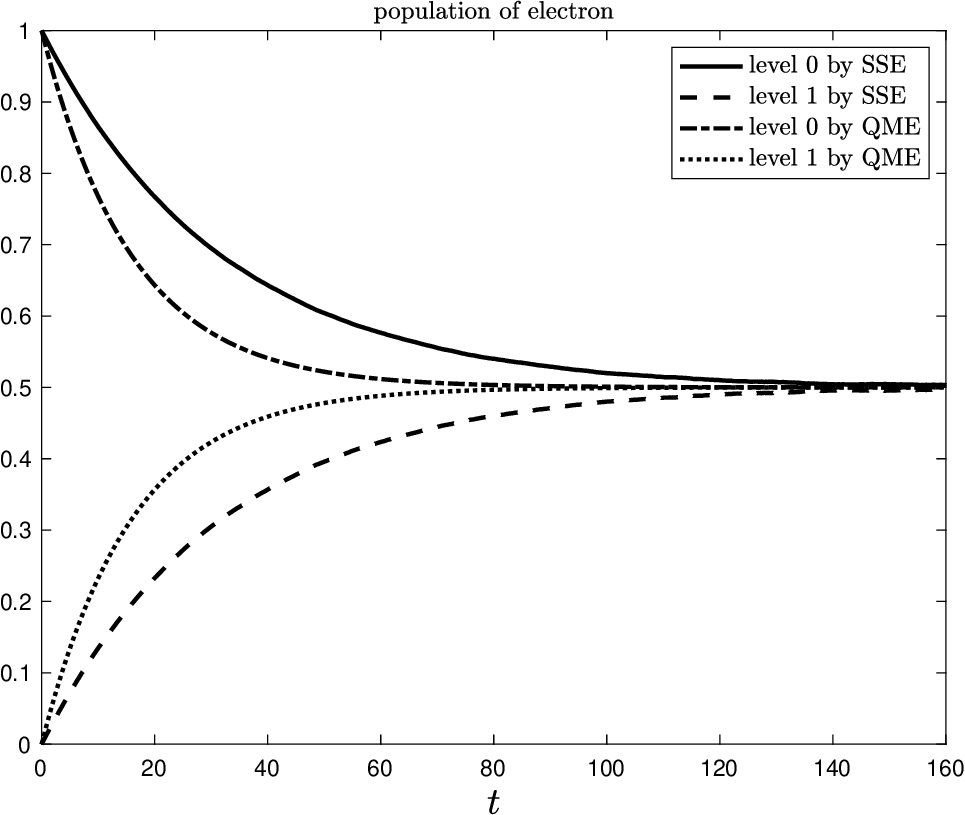}}
    \caption{The population of electrons on two levels corresponding to $\varepsilon=1/32$. From the top left to the bottom right, $\lambda$ corresponds to $\varepsilon,\varepsilon/2,\varepsilon/4$, and $\varepsilon/8$, respectively. The sample size of SSE is $M=80000$.}
    \label{fig: sse_vs_qme_1}
\end{figure*}

\begin{figure*}[htbp]
    \centering
    \subfigbottomskip=2pt 
    \subfigcapskip=-5pt 
    \subfigure{
		\includegraphics[width=0.23\linewidth]{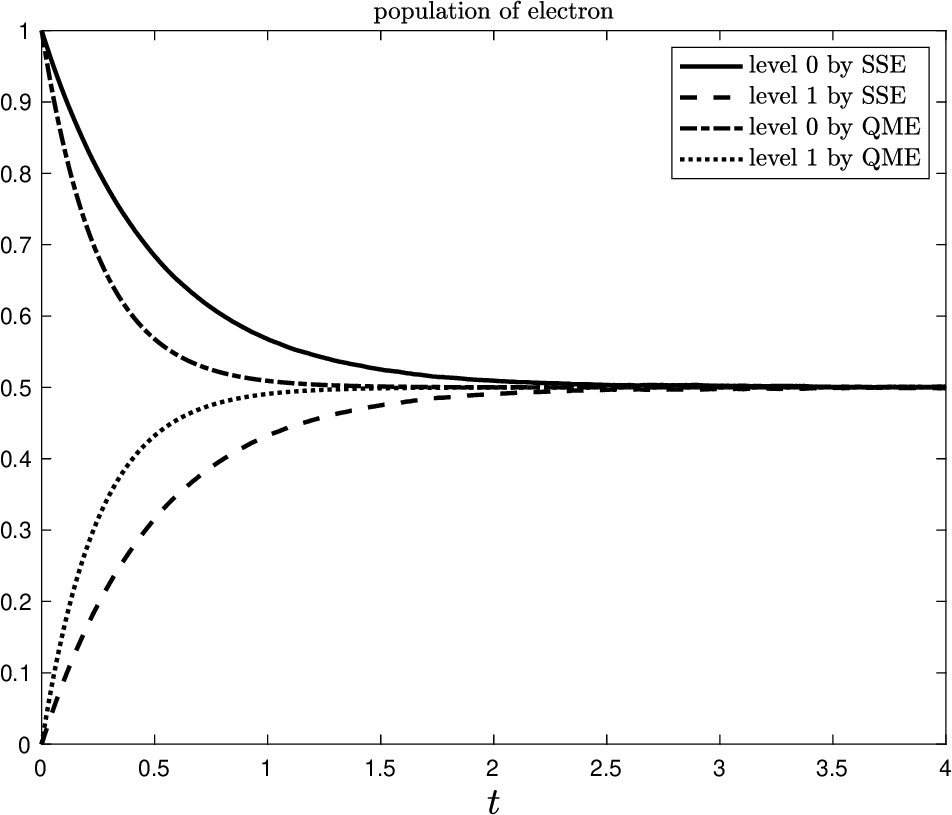}}
    \subfigure{
		\includegraphics[width=0.23\linewidth]{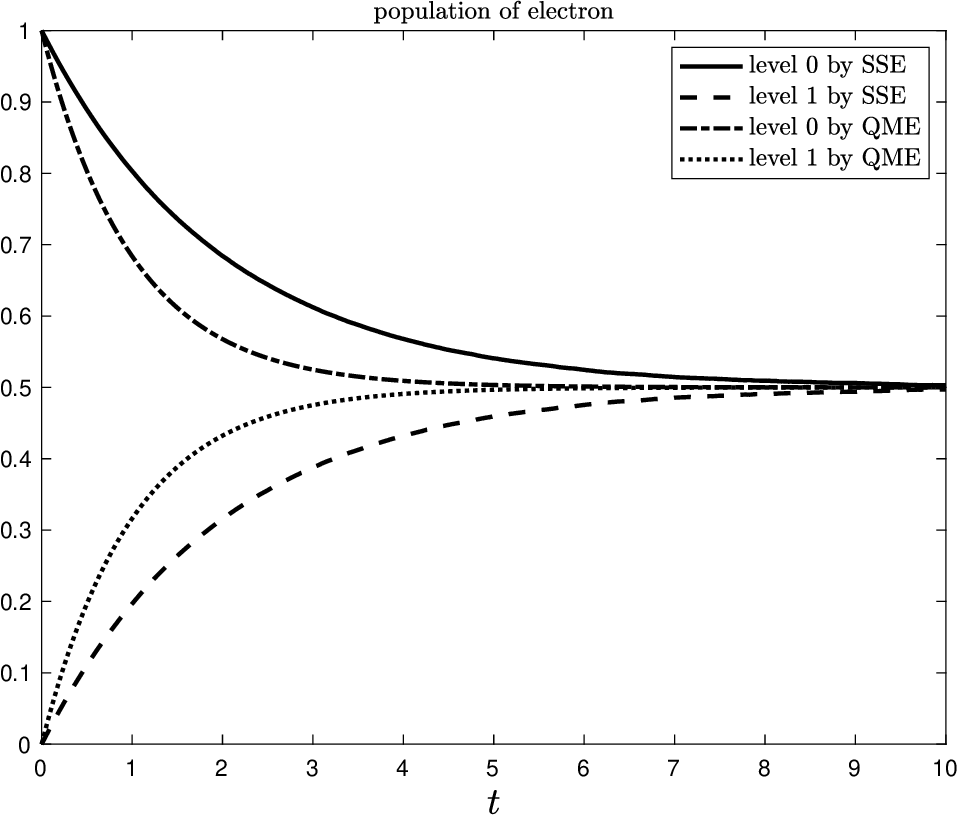}}
    \subfigure{
		\includegraphics[width=0.23\linewidth]{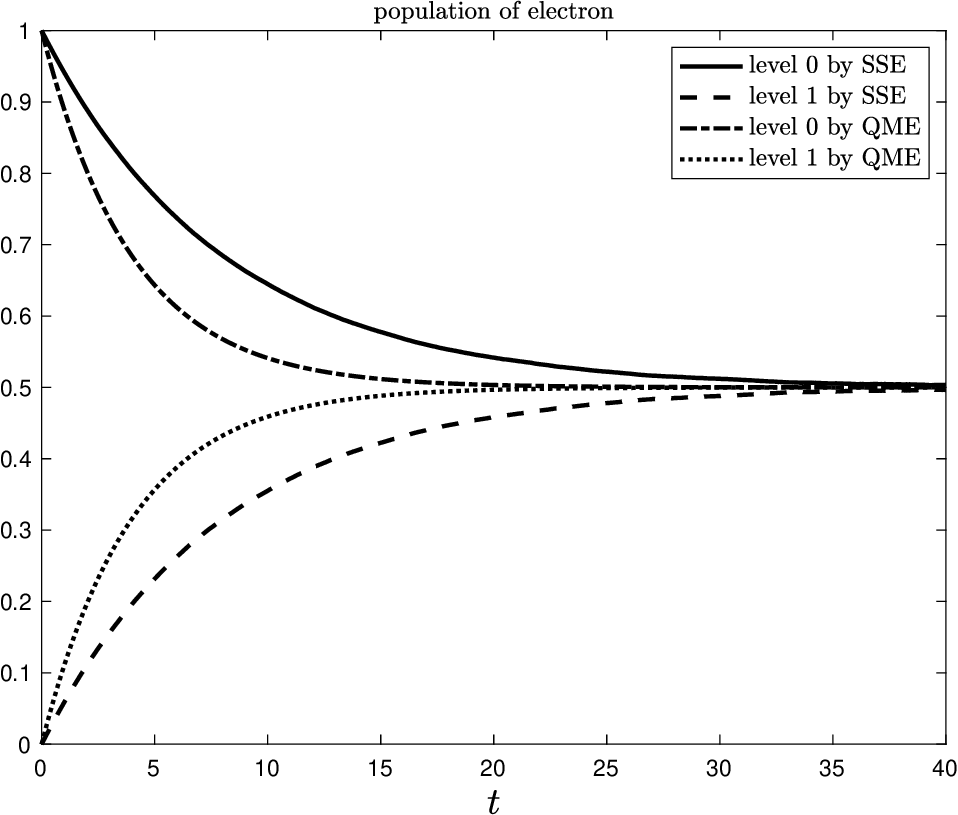}}
    \subfigure{
		\includegraphics[width=0.23\linewidth]{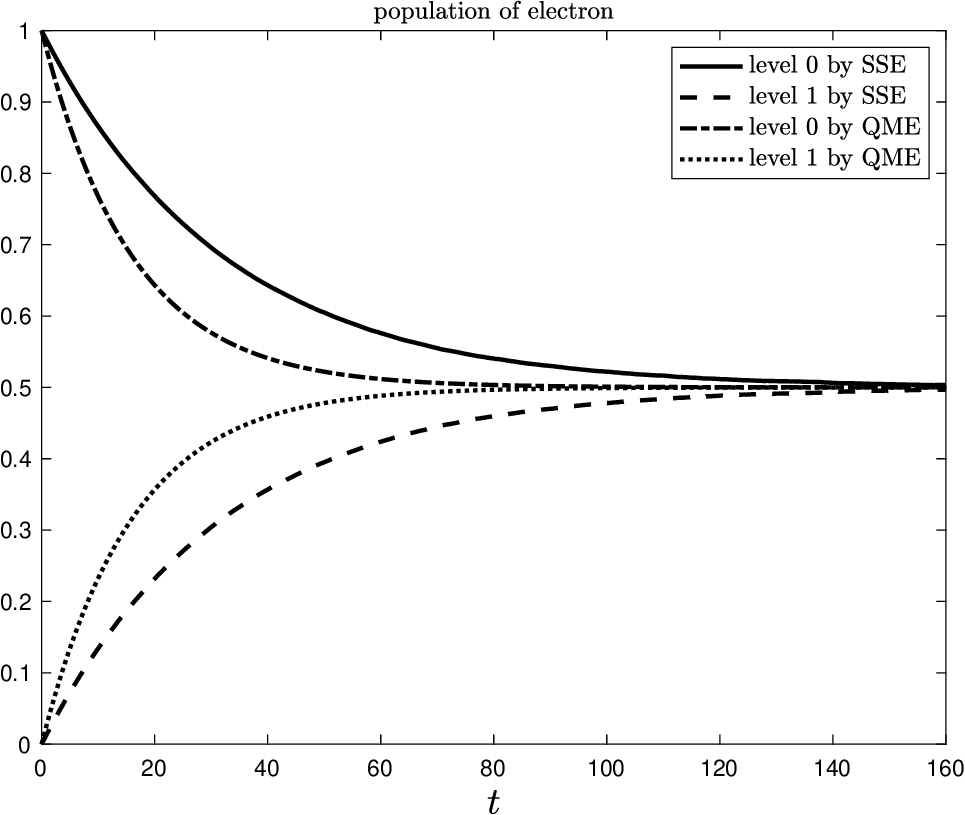}}
    \caption{The population of electrons on two levels corresponding to $\varepsilon=1/64$. From the top left to the bottom right, $\lambda$ corresponds to $\varepsilon,\varepsilon/2,\varepsilon/4$, and $\varepsilon/8$, respectively. The sample size of SSE is $M=80000$.}
    \label{fig: sse_vs_qme_2}
\end{figure*}

In the two numerical experiments, we test with two different $\varepsilon=1/32$ and $1/64$. For each $\varepsilon$, we let the $\lambda$ vary from $\varepsilon$ to $\varepsilon/8$.
The evolutions of the population of electrons on two levels based on SSE and QME for different $\varepsilon$ and $\lambda$ are shown in \cref{fig: sse_vs_qme_1} and \cref{fig: sse_vs_qme_2}. The result indicates that (1) in this parameter setting, SSE and QME have the same equilibrium but different transient dynamics, and QME will reach the steady state faster than SSE; (2) for a fixed $\varepsilon$, $\lambda$ only affects the relaxation time to reach the equilibrium,  and a smaller $\lambda$ indicates a longer relaxation time. 

The former observation can be understood in two ways. Mathematically, the QME is not exactly equivalent to SSE, they are equal only if the higher-order term $O(\lambda^3)$ is ignored. In a short time, the higher order terms lead to a difference in the dynamical behavior of the two, but in a long time, the higher order terms cancel each other out and thus have no effect on the steady state. Scientifically, QME is averaged over the stochastic term, such that the dissipation terms do not compete with the random fluctuations in dynamics, and thus the dynamics corresponding to QME will reach the steady state faster. This is a common relationship between SSE and QME and has been discussed in \cite{biele2012stochastic}. The latter shows that SSE still has the same steady state as QME when $\lambda$ is not much less than $\varepsilon$, at least for a large class of parameter settings. This indicates that $\lambda\ll\varepsilon$ is only a sufficient condition, not a necessary one to establish qualitative connections between SSE and QME. More comparison studies and even qualitative connections are worth exploring in the future.

\section{Conclusion and Discussions}
In this article, we propose a stochastic Schr\"{o}dinger equation model for the Anderson-Holstein impurities, filling a gap in the study of open quantum systems with fermionic baths. The SSE model is obtained directly from the microscopic model instead of using empirical correlation functions. Through analytic derivations, we establish the theoretical relations between AH, SSE and QME, and introduce the hierarchy of modeling in \cref{fig:nMSSE}. We also discuss efficient algorithms for noise generation and sampling stochastic trajectories. Our numerical experiments show that SSE could be used to study physical observables and capture effects beyond the level of QME.

From the computational perspective, the non-Markovian term could be potentially expensive, especially in the high-dimensional case.  If one is interested in the nonequilibrium dynamics of SSE, efficient algorithms that incorporate noise and treat the non-Markovian term with (quasi-)linear cost in time (for example, see \cite{kaye2023fast}) would be required to propagate towards a longer time.

Our current studies focus on Holstein types of quantum impurities. Similar strategies are applicable to correlated systems via the pseudoparticle approach \cite{eckstein2010nonequilibrium}. Understanding decoherence and relaxation dynamics of interacting systems \cite{haertle2013decoherence,haertle2014formation} will be the focus of our future work.

\text{ }

\section*{Acknowledgement}

Z.Z. is supported by the National Key R\&D Program of China, Project Number 2021YFA1001200, and the NSFC, grant Number 12031013, 12171013. This work is partially supported by a grant from the Simons Foundation under Award No. 825053 (Z.H.).
We thank helpful discussions with Dr. Yu Cao. L.X. thanks Dr. Hao Wu for his support and encouragement.

All authors contribute equally to this work. 

\bibliographystyle{apsrev}
\bibliography{sse}

\begin{thebibliography}{57}
\expandafter\ifx\csname natexlab\endcsname\relax\def\natexlab#1{#1}\fi
\expandafter\ifx\csname bibnamefont\endcsname\relax
  \def\bibnamefont#1{#1}\fi
\expandafter\ifx\csname bibfnamefont\endcsname\relax
  \def\bibfnamefont#1{#1}\fi
\expandafter\ifx\csname citenamefont\endcsname\relax
  \def\citenamefont#1{#1}\fi
\expandafter\ifx\csname url\endcsname\relax
  \def\url#1{\texttt{#1}}\fi
\expandafter\ifx\csname urlprefix\endcsname\relax\def\urlprefix{URL }\fi
\providecommand{\bibinfo}[2]{#2}
\providecommand{\eprint}[2][]{\url{#2}}

\bibitem[{\citenamefont{Anderson}(1961)}]{anderson1961localized}
\bibinfo{author}{\bibfnamefont{P.~W.} \bibnamefont{Anderson}},
  \bibinfo{journal}{Physical Review} \textbf{\bibinfo{volume}{124}},
  \bibinfo{pages}{41} (\bibinfo{year}{1961}).

\bibitem[{\citenamefont{Hanson et~al.}(2007)\citenamefont{Hanson, Kouwenhoven,
  Petta, Tarucha, and Vandersypen}}]{hanson2007spins}
\bibinfo{author}{\bibfnamefont{R.}~\bibnamefont{Hanson}},
  \bibinfo{author}{\bibfnamefont{L.~P.} \bibnamefont{Kouwenhoven}},
  \bibinfo{author}{\bibfnamefont{J.~R.} \bibnamefont{Petta}},
  \bibinfo{author}{\bibfnamefont{S.}~\bibnamefont{Tarucha}}, \bibnamefont{and}
  \bibinfo{author}{\bibfnamefont{L.~M.} \bibnamefont{Vandersypen}},
  \bibinfo{journal}{Reviews of modern physics} \textbf{\bibinfo{volume}{79}},
  \bibinfo{pages}{1217} (\bibinfo{year}{2007}).

\bibitem[{\citenamefont{Brako and Newns}(1981)}]{brako1981slowly}
\bibinfo{author}{\bibfnamefont{R.}~\bibnamefont{Brako}} \bibnamefont{and}
  \bibinfo{author}{\bibfnamefont{D.}~\bibnamefont{Newns}},
  \bibinfo{journal}{Journal of Physics C: Solid State Physics}
  \textbf{\bibinfo{volume}{14}}, \bibinfo{pages}{3065} (\bibinfo{year}{1981}).

\bibitem[{\citenamefont{Newns}(1969)}]{newns1969self}
\bibinfo{author}{\bibfnamefont{D.}~\bibnamefont{Newns}},
  \bibinfo{journal}{Physical Review} \textbf{\bibinfo{volume}{178}},
  \bibinfo{pages}{1123} (\bibinfo{year}{1969}).

\bibitem[{\citenamefont{Holstein}(1959)}]{holstein1959studies}
\bibinfo{author}{\bibfnamefont{T.}~\bibnamefont{Holstein}},
  \bibinfo{journal}{Annals of physics} \textbf{\bibinfo{volume}{8}},
  \bibinfo{pages}{325} (\bibinfo{year}{1959}).

\bibitem[{\citenamefont{Persson}(1993)}]{persson1993applications}
\bibinfo{author}{\bibfnamefont{B.}~\bibnamefont{Persson}},
  \bibinfo{journal}{The Journal of Chemical Physics}
  \textbf{\bibinfo{volume}{98}}, \bibinfo{pages}{1659} (\bibinfo{year}{1993}).

\bibitem[{\citenamefont{Luo et~al.}(2016)\citenamefont{Luo, Jiang, Juaristi,
  Alducin, and Guo}}]{luo2016electron}
\bibinfo{author}{\bibfnamefont{X.}~\bibnamefont{Luo}},
  \bibinfo{author}{\bibfnamefont{B.}~\bibnamefont{Jiang}},
  \bibinfo{author}{\bibfnamefont{J.~I.} \bibnamefont{Juaristi}},
  \bibinfo{author}{\bibfnamefont{M.}~\bibnamefont{Alducin}}, \bibnamefont{and}
  \bibinfo{author}{\bibfnamefont{H.}~\bibnamefont{Guo}}, \bibinfo{journal}{The
  Journal of Chemical Physics} \textbf{\bibinfo{volume}{145}},
  \bibinfo{pages}{044704} (\bibinfo{year}{2016}).

\bibitem[{\citenamefont{Nitzan and Ratner}(2003)}]{nitzan2003electron}
\bibinfo{author}{\bibfnamefont{A.}~\bibnamefont{Nitzan}} \bibnamefont{and}
  \bibinfo{author}{\bibfnamefont{M.~A.} \bibnamefont{Ratner}},
  \bibinfo{journal}{Science} \textbf{\bibinfo{volume}{300}},
  \bibinfo{pages}{1384} (\bibinfo{year}{2003}).

\bibitem[{\citenamefont{Shenvi et~al.}(2009)\citenamefont{Shenvi, Roy, and
  Tully}}]{shenvi2009nonadiabatic}
\bibinfo{author}{\bibfnamefont{N.}~\bibnamefont{Shenvi}},
  \bibinfo{author}{\bibfnamefont{S.}~\bibnamefont{Roy}}, \bibnamefont{and}
  \bibinfo{author}{\bibfnamefont{J.~C.} \bibnamefont{Tully}},
  \bibinfo{journal}{The Journal of chemical physics}
  \textbf{\bibinfo{volume}{130}}, \bibinfo{pages}{174107}
  (\bibinfo{year}{2009}).

\bibitem[{\citenamefont{Huang et~al.}(2023{\natexlab{a}})\citenamefont{Huang,
  Xu, and Zhou}}]{huang2023efficient}
\bibinfo{author}{\bibfnamefont{Z.}~\bibnamefont{Huang}},
  \bibinfo{author}{\bibfnamefont{L.}~\bibnamefont{Xu}}, \bibnamefont{and}
  \bibinfo{author}{\bibfnamefont{Z.}~\bibnamefont{Zhou}},
  \bibinfo{journal}{Journal of Computational Physics}
  \textbf{\bibinfo{volume}{474}}, \bibinfo{pages}{111771}
  (\bibinfo{year}{2023}{\natexlab{a}}).

\bibitem[{\citenamefont{Katz et~al.}(2005)\citenamefont{Katz, Zeiri, and
  Kosloff}}]{katz2005role}
\bibinfo{author}{\bibfnamefont{G.}~\bibnamefont{Katz}},
  \bibinfo{author}{\bibfnamefont{Y.}~\bibnamefont{Zeiri}}, \bibnamefont{and}
  \bibinfo{author}{\bibfnamefont{R.}~\bibnamefont{Kosloff}},
  \bibinfo{journal}{The Journal of Physical Chemistry B}
  \textbf{\bibinfo{volume}{109}}, \bibinfo{pages}{18876}
  (\bibinfo{year}{2005}).

\bibitem[{\citenamefont{Shenvi et~al.}(2006)\citenamefont{Shenvi, Roy,
  Parandekar, and Tully}}]{shenvi2006vibrational}
\bibinfo{author}{\bibfnamefont{N.}~\bibnamefont{Shenvi}},
  \bibinfo{author}{\bibfnamefont{S.}~\bibnamefont{Roy}},
  \bibinfo{author}{\bibfnamefont{P.}~\bibnamefont{Parandekar}},
  \bibnamefont{and} \bibinfo{author}{\bibfnamefont{J.}~\bibnamefont{Tully}},
  \bibinfo{journal}{The Journal of chemical physics}
  \textbf{\bibinfo{volume}{125}} (\bibinfo{year}{2006}).

\bibitem[{\citenamefont{Langreth and
  Nordlander}(1991)}]{langreth1991derivation}
\bibinfo{author}{\bibfnamefont{D.~C.} \bibnamefont{Langreth}} \bibnamefont{and}
  \bibinfo{author}{\bibfnamefont{P.}~\bibnamefont{Nordlander}},
  \bibinfo{journal}{Physical Review B} \textbf{\bibinfo{volume}{43}},
  \bibinfo{pages}{2541} (\bibinfo{year}{1991}).

\bibitem[{\citenamefont{Dou and Subotnik}(2016)}]{dou2016broadened}
\bibinfo{author}{\bibfnamefont{W.}~\bibnamefont{Dou}} \bibnamefont{and}
  \bibinfo{author}{\bibfnamefont{J.~E.} \bibnamefont{Subotnik}},
  \bibinfo{journal}{The Journal of chemical physics}
  \textbf{\bibinfo{volume}{144}}, \bibinfo{pages}{024116}
  (\bibinfo{year}{2016}).

\bibitem[{\citenamefont{Ouyang et~al.}(2015)\citenamefont{Ouyang, Saven, and
  Subotnik}}]{ouyang2015surface}
\bibinfo{author}{\bibfnamefont{W.}~\bibnamefont{Ouyang}},
  \bibinfo{author}{\bibfnamefont{J.~G.} \bibnamefont{Saven}}, \bibnamefont{and}
  \bibinfo{author}{\bibfnamefont{J.~E.} \bibnamefont{Subotnik}},
  \bibinfo{journal}{The Journal of Physical Chemistry C}
  \textbf{\bibinfo{volume}{119}}, \bibinfo{pages}{20833}
  (\bibinfo{year}{2015}).

\bibitem[{\citenamefont{Miao et~al.}(2019)\citenamefont{Miao, Ouyang, and
  Subotnik}}]{miao2019comparison}
\bibinfo{author}{\bibfnamefont{G.}~\bibnamefont{Miao}},
  \bibinfo{author}{\bibfnamefont{W.}~\bibnamefont{Ouyang}}, \bibnamefont{and}
  \bibinfo{author}{\bibfnamefont{J.}~\bibnamefont{Subotnik}},
  \bibinfo{journal}{The Journal of Chemical Physics}
  \textbf{\bibinfo{volume}{150}} (\bibinfo{year}{2019}).

\bibitem[{\citenamefont{Gao}(1997)}]{gao1997dissipative}
\bibinfo{author}{\bibfnamefont{S.}~\bibnamefont{Gao}},
  \bibinfo{journal}{Physical review letters} \textbf{\bibinfo{volume}{79}},
  \bibinfo{pages}{3101} (\bibinfo{year}{1997}).

\bibitem[{\citenamefont{Wang et~al.}(2020)\citenamefont{Wang, Nijjar, Zhou,
  Bondar, and Prezhdo}}]{wang2020combining}
\bibinfo{author}{\bibfnamefont{Y.-S.} \bibnamefont{Wang}},
  \bibinfo{author}{\bibfnamefont{P.}~\bibnamefont{Nijjar}},
  \bibinfo{author}{\bibfnamefont{X.}~\bibnamefont{Zhou}},
  \bibinfo{author}{\bibfnamefont{D.~I.} \bibnamefont{Bondar}},
  \bibnamefont{and} \bibinfo{author}{\bibfnamefont{O.~V.}
  \bibnamefont{Prezhdo}}, \bibinfo{journal}{The Journal of Physical Chemistry
  B} \textbf{\bibinfo{volume}{124}}, \bibinfo{pages}{4326}
  (\bibinfo{year}{2020}).

\bibitem[{\citenamefont{Nakajima}(1958)}]{nakajima1958quantum}
\bibinfo{author}{\bibfnamefont{S.}~\bibnamefont{Nakajima}},
  \bibinfo{journal}{Progress of Theoretical Physics}
  \textbf{\bibinfo{volume}{20}}, \bibinfo{pages}{948} (\bibinfo{year}{1958}).

\bibitem[{\citenamefont{Fay}(2021)}]{fay2021chirality}
\bibinfo{author}{\bibfnamefont{T.~P.} \bibnamefont{Fay}}, \bibinfo{journal}{The
  Journal of Physical Chemistry Letters} \textbf{\bibinfo{volume}{12}},
  \bibinfo{pages}{1407} (\bibinfo{year}{2021}).

\bibitem[{\citenamefont{Redfield}(1965)}]{redfield1965theory}
\bibinfo{author}{\bibfnamefont{A.~G.} \bibnamefont{Redfield}}, in
  \emph{\bibinfo{booktitle}{Advances in Magnetic and Optical Resonance}}
  (\bibinfo{publisher}{Elsevier}, \bibinfo{year}{1965}),
  vol.~\bibinfo{volume}{1}, pp. \bibinfo{pages}{1--32}.

\bibitem[{\citenamefont{Leathers and Micha}(2006)}]{leathers2006density}
\bibinfo{author}{\bibfnamefont{A.~S.} \bibnamefont{Leathers}} \bibnamefont{and}
  \bibinfo{author}{\bibfnamefont{D.~A.} \bibnamefont{Micha}},
  \bibinfo{journal}{The Journal of Physical Chemistry A}
  \textbf{\bibinfo{volume}{110}}, \bibinfo{pages}{749} (\bibinfo{year}{2006}).

\bibitem[{\citenamefont{Elste et~al.}(2008)\citenamefont{Elste, Weick, Timm,
  and von Oppen}}]{elste2008current}
\bibinfo{author}{\bibfnamefont{F.}~\bibnamefont{Elste}},
  \bibinfo{author}{\bibfnamefont{G.}~\bibnamefont{Weick}},
  \bibinfo{author}{\bibfnamefont{C.}~\bibnamefont{Timm}}, \bibnamefont{and}
  \bibinfo{author}{\bibfnamefont{F.}~\bibnamefont{von Oppen}},
  \bibinfo{journal}{Applied Physics A} \textbf{\bibinfo{volume}{93}},
  \bibinfo{pages}{345} (\bibinfo{year}{2008}).

\bibitem[{\citenamefont{Elste et~al.}(2010)\citenamefont{Elste, Reichman, and
  Millis}}]{elste2010effect}
\bibinfo{author}{\bibfnamefont{F.}~\bibnamefont{Elste}},
  \bibinfo{author}{\bibfnamefont{D.~R.} \bibnamefont{Reichman}},
  \bibnamefont{and} \bibinfo{author}{\bibfnamefont{A.~J.}
  \bibnamefont{Millis}}, \bibinfo{journal}{Physical Review B}
  \textbf{\bibinfo{volume}{81}}, \bibinfo{pages}{205413}
  (\bibinfo{year}{2010}).

\bibitem[{\citenamefont{Elste et~al.}(2011{\natexlab{a}})\citenamefont{Elste,
  Reichman, and Millis}}]{elste2011transport}
\bibinfo{author}{\bibfnamefont{F.}~\bibnamefont{Elste}},
  \bibinfo{author}{\bibfnamefont{D.~R.} \bibnamefont{Reichman}},
  \bibnamefont{and} \bibinfo{author}{\bibfnamefont{A.~J.}
  \bibnamefont{Millis}}, \bibinfo{journal}{Physical Review B}
  \textbf{\bibinfo{volume}{83}}, \bibinfo{pages}{085415}
  (\bibinfo{year}{2011}{\natexlab{a}}).

\bibitem[{\citenamefont{Elste et~al.}(2011{\natexlab{b}})\citenamefont{Elste,
  Reichman, and Millis}}]{elste2011transport2}
\bibinfo{author}{\bibfnamefont{F.}~\bibnamefont{Elste}},
  \bibinfo{author}{\bibfnamefont{D.~R.} \bibnamefont{Reichman}},
  \bibnamefont{and} \bibinfo{author}{\bibfnamefont{A.~J.}
  \bibnamefont{Millis}}, \bibinfo{journal}{Physical Review B}
  \textbf{\bibinfo{volume}{83}}, \bibinfo{pages}{245405}
  (\bibinfo{year}{2011}{\natexlab{b}}).

\bibitem[{\citenamefont{Haertle et~al.}(2013)\citenamefont{Haertle, Cohen,
  Reichman, and Millis}}]{haertle2013decoherence}
\bibinfo{author}{\bibfnamefont{R.}~\bibnamefont{Haertle}},
  \bibinfo{author}{\bibfnamefont{G.}~\bibnamefont{Cohen}},
  \bibinfo{author}{\bibfnamefont{D.}~\bibnamefont{Reichman}}, \bibnamefont{and}
  \bibinfo{author}{\bibfnamefont{A.}~\bibnamefont{Millis}},
  \bibinfo{journal}{Physical Review B} \textbf{\bibinfo{volume}{88}},
  \bibinfo{pages}{235426} (\bibinfo{year}{2013}).

\bibitem[{\citenamefont{Tanimura and Kubo}(1989)}]{tanimura1989time}
\bibinfo{author}{\bibfnamefont{Y.}~\bibnamefont{Tanimura}} \bibnamefont{and}
  \bibinfo{author}{\bibfnamefont{R.}~\bibnamefont{Kubo}},
  \bibinfo{journal}{Journal of the Physical Society of Japan}
  \textbf{\bibinfo{volume}{58}}, \bibinfo{pages}{101} (\bibinfo{year}{1989}).

\bibitem[{\citenamefont{Erpenbeck and Thoss}(2019)}]{erpenbeck2019hierarchical}
\bibinfo{author}{\bibfnamefont{A.}~\bibnamefont{Erpenbeck}} \bibnamefont{and}
  \bibinfo{author}{\bibfnamefont{M.}~\bibnamefont{Thoss}},
  \bibinfo{journal}{The Journal of Chemical Physics}
  \textbf{\bibinfo{volume}{151}} (\bibinfo{year}{2019}).

\bibitem[{\citenamefont{Goetsch et~al.}(1996)\citenamefont{Goetsch, Tombesi,
  and Vitali}}]{goetsch1996effect}
\bibinfo{author}{\bibfnamefont{P.}~\bibnamefont{Goetsch}},
  \bibinfo{author}{\bibfnamefont{P.}~\bibnamefont{Tombesi}}, \bibnamefont{and}
  \bibinfo{author}{\bibfnamefont{D.}~\bibnamefont{Vitali}},
  \bibinfo{journal}{Physical Review A} \textbf{\bibinfo{volume}{54}},
  \bibinfo{pages}{4519} (\bibinfo{year}{1996}).

\bibitem[{\citenamefont{Shiokawa and Hu}(1995)}]{shiokawa1995decoherence}
\bibinfo{author}{\bibfnamefont{K.}~\bibnamefont{Shiokawa}} \bibnamefont{and}
  \bibinfo{author}{\bibfnamefont{B.}~\bibnamefont{Hu}},
  \bibinfo{journal}{Physical Review E} \textbf{\bibinfo{volume}{52}},
  \bibinfo{pages}{2497} (\bibinfo{year}{1995}).

\bibitem[{\citenamefont{Wiseman}(1996)}]{wiseman1996quantum}
\bibinfo{author}{\bibfnamefont{H.~M.} \bibnamefont{Wiseman}},
  \bibinfo{journal}{Quantum and Semiclassical Optics: Journal of the European
  Optical Society Part B} \textbf{\bibinfo{volume}{8}}, \bibinfo{pages}{205}
  (\bibinfo{year}{1996}).

\bibitem[{\citenamefont{Gambetta and Wiseman}(2002)}]{gambetta2002non}
\bibinfo{author}{\bibfnamefont{J.}~\bibnamefont{Gambetta}} \bibnamefont{and}
  \bibinfo{author}{\bibfnamefont{H.~M.} \bibnamefont{Wiseman}},
  \bibinfo{journal}{Physical Review A} \textbf{\bibinfo{volume}{66}},
  \bibinfo{pages}{012108} (\bibinfo{year}{2002}).

\bibitem[{\citenamefont{Gambetta and
  Wiseman}(2003)}]{gambetta2003interpretation}
\bibinfo{author}{\bibfnamefont{J.}~\bibnamefont{Gambetta}} \bibnamefont{and}
  \bibinfo{author}{\bibfnamefont{H.}~\bibnamefont{Wiseman}},
  \bibinfo{journal}{Physical Review A} \textbf{\bibinfo{volume}{68}},
  \bibinfo{pages}{062104} (\bibinfo{year}{2003}).

\bibitem[{\citenamefont{Breuer and Petruccione}(1995)}]{breuer1995stochastic}
\bibinfo{author}{\bibfnamefont{H.-P.} \bibnamefont{Breuer}} \bibnamefont{and}
  \bibinfo{author}{\bibfnamefont{F.}~\bibnamefont{Petruccione}},
  \bibinfo{journal}{Physical Review E} \textbf{\bibinfo{volume}{52}},
  \bibinfo{pages}{428} (\bibinfo{year}{1995}).

\bibitem[{\citenamefont{Breuer and Piilo}(2009)}]{breuer2009stochastic}
\bibinfo{author}{\bibfnamefont{H.-P.} \bibnamefont{Breuer}} \bibnamefont{and}
  \bibinfo{author}{\bibfnamefont{J.}~\bibnamefont{Piilo}},
  \bibinfo{journal}{Europhysics Letters} \textbf{\bibinfo{volume}{85}},
  \bibinfo{pages}{50004} (\bibinfo{year}{2009}).

\bibitem[{\citenamefont{Zhao et~al.}(2012)\citenamefont{Zhao, Shi, Wu, and
  Yu}}]{zhao2012fermionic}
\bibinfo{author}{\bibfnamefont{X.}~\bibnamefont{Zhao}},
  \bibinfo{author}{\bibfnamefont{W.}~\bibnamefont{Shi}},
  \bibinfo{author}{\bibfnamefont{L.-A.} \bibnamefont{Wu}}, \bibnamefont{and}
  \bibinfo{author}{\bibfnamefont{T.}~\bibnamefont{Yu}},
  \bibinfo{journal}{Physical Review A} \textbf{\bibinfo{volume}{86}},
  \bibinfo{pages}{032116} (\bibinfo{year}{2012}).

\bibitem[{\citenamefont{Nathan and Rudner}(2020)}]{nathan2020universal}
\bibinfo{author}{\bibfnamefont{F.}~\bibnamefont{Nathan}} \bibnamefont{and}
  \bibinfo{author}{\bibfnamefont{M.~S.} \bibnamefont{Rudner}},
  \bibinfo{journal}{Physical Review B} \textbf{\bibinfo{volume}{102}},
  \bibinfo{pages}{115109} (\bibinfo{year}{2020}).

\bibitem[{\citenamefont{Breuer et~al.}(2002)\citenamefont{Breuer, Petruccione
  et~al.}}]{breuer2002theory}
\bibinfo{author}{\bibfnamefont{H.-P.} \bibnamefont{Breuer}},
  \bibinfo{author}{\bibfnamefont{F.}~\bibnamefont{Petruccione}},
  \bibnamefont{et~al.}, \emph{\bibinfo{title}{The theory of open quantum
  systems}} (\bibinfo{publisher}{Oxford University Press on Demand},
  \bibinfo{year}{2002}).

\bibitem[{\citenamefont{Strunz and Yu}(2004)}]{strunz2004convolutionless}
\bibinfo{author}{\bibfnamefont{W.~T.} \bibnamefont{Strunz}} \bibnamefont{and}
  \bibinfo{author}{\bibfnamefont{T.}~\bibnamefont{Yu}},
  \bibinfo{journal}{Physical Review A} \textbf{\bibinfo{volume}{69}},
  \bibinfo{pages}{052115} (\bibinfo{year}{2004}).

\bibitem[{\citenamefont{Jentzen and Kloeden}(2011)}]{jentzen2011taylor}
\bibinfo{author}{\bibfnamefont{A.}~\bibnamefont{Jentzen}} \bibnamefont{and}
  \bibinfo{author}{\bibfnamefont{P.~E.} \bibnamefont{Kloeden}},
  \emph{\bibinfo{title}{Taylor approximations for stochastic partial
  differential equations}} (\bibinfo{publisher}{SIAM}, \bibinfo{year}{2011}).

\bibitem[{\citenamefont{Zhang and Karniadakis}(2017)}]{zhang2017numerical}
\bibinfo{author}{\bibfnamefont{Z.}~\bibnamefont{Zhang}} \bibnamefont{and}
  \bibinfo{author}{\bibfnamefont{G.}~\bibnamefont{Karniadakis}},
  \emph{\bibinfo{title}{Numerical methods for stochastic partial differential
  equations with white noise}}, vol. \bibinfo{volume}{196}
  (\bibinfo{publisher}{Springer}, \bibinfo{year}{2017}).

\bibitem[{\citenamefont{Shi et~al.}(2013)\citenamefont{Shi, Zhao, and
  Yu}}]{shi2013non}
\bibinfo{author}{\bibfnamefont{W.}~\bibnamefont{Shi}},
  \bibinfo{author}{\bibfnamefont{X.}~\bibnamefont{Zhao}}, \bibnamefont{and}
  \bibinfo{author}{\bibfnamefont{T.}~\bibnamefont{Yu}},
  \bibinfo{journal}{Physical Review A} \textbf{\bibinfo{volume}{87}},
  \bibinfo{pages}{052127} (\bibinfo{year}{2013}).

\bibitem[{\citenamefont{Cao and Lu}(2017)}]{cao_lindblad_2017}
\bibinfo{author}{\bibfnamefont{Y.}~\bibnamefont{Cao}} \bibnamefont{and}
  \bibinfo{author}{\bibfnamefont{J.}~\bibnamefont{Lu}},
  \bibinfo{journal}{Journal of Mathematical Physics}
  \textbf{\bibinfo{volume}{58}}, \bibinfo{pages}{122105}
  (\bibinfo{year}{2017}), ISSN \bibinfo{issn}{0022-2488, 1089-7658},
  \urlprefix\url{http://aip.scitation.org/doi/10.1063/1.4993431}.

\bibitem[{\citenamefont{Bogoliubov}(1947)}]{bogoliubov1947theory}
\bibinfo{author}{\bibfnamefont{N.}~\bibnamefont{Bogoliubov}},
  \bibinfo{journal}{J. Phys} \textbf{\bibinfo{volume}{11}}, \bibinfo{pages}{23}
  (\bibinfo{year}{1947}).

\bibitem[{\citenamefont{Bogoljubov et~al.}(1958)\citenamefont{Bogoljubov,
  Tolmachov, and {\v{S}}irkov}}]{bogoljubov1958new}
\bibinfo{author}{\bibfnamefont{N.}~\bibnamefont{Bogoljubov}},
  \bibinfo{author}{\bibfnamefont{V.~V.} \bibnamefont{Tolmachov}},
  \bibnamefont{and}
  \bibinfo{author}{\bibfnamefont{D.}~\bibnamefont{{\v{S}}irkov}},
  \bibinfo{journal}{Fortschritte der physik} \textbf{\bibinfo{volume}{6}},
  \bibinfo{pages}{605} (\bibinfo{year}{1958}).

\bibitem[{\citenamefont{Gaspard and Nagaoka}(1999)}]{gaspard1999non}
\bibinfo{author}{\bibfnamefont{P.}~\bibnamefont{Gaspard}} \bibnamefont{and}
  \bibinfo{author}{\bibfnamefont{M.}~\bibnamefont{Nagaoka}},
  \bibinfo{journal}{The Journal of chemical physics}
  \textbf{\bibinfo{volume}{111}}, \bibinfo{pages}{5676} (\bibinfo{year}{1999}).

\bibitem[{\citenamefont{Dou et~al.}(2015)\citenamefont{Dou, Nitzan, and
  Subotnik}}]{dou_surface_2015}
\bibinfo{author}{\bibfnamefont{W.}~\bibnamefont{Dou}},
  \bibinfo{author}{\bibfnamefont{A.}~\bibnamefont{Nitzan}}, \bibnamefont{and}
  \bibinfo{author}{\bibfnamefont{J.~E.} \bibnamefont{Subotnik}},
  \bibinfo{journal}{The Journal of Chemical Physics}
  \textbf{\bibinfo{volume}{142}}, \bibinfo{pages}{084110}
  (\bibinfo{year}{2015}), ISSN \bibinfo{issn}{0021-9606, 1089-7690},
  \urlprefix\url{http://aip.scitation.org/doi/10.1063/1.4908034}.

\bibitem[{\citenamefont{Dou et~al.}(2017)\citenamefont{Dou, Miao, and
  Subotnik}}]{dou2017born}
\bibinfo{author}{\bibfnamefont{W.}~\bibnamefont{Dou}},
  \bibinfo{author}{\bibfnamefont{G.}~\bibnamefont{Miao}}, \bibnamefont{and}
  \bibinfo{author}{\bibfnamefont{J.~E.} \bibnamefont{Subotnik}},
  \bibinfo{journal}{Physical Review Letters} \textbf{\bibinfo{volume}{119}},
  \bibinfo{pages}{046001} (\bibinfo{year}{2017}).

\bibitem[{\citenamefont{Trefethen}(2019)}]{trefethen2019approximation}
\bibinfo{author}{\bibfnamefont{L.~N.} \bibnamefont{Trefethen}},
  \emph{\bibinfo{title}{Approximation Theory and Approximation Practice,
  Extended Edition}} (\bibinfo{publisher}{SIAM}, \bibinfo{year}{2019}).

\bibitem[{\citenamefont{Huang et~al.}(2023{\natexlab{b}})\citenamefont{Huang,
  Gull, and Lin}}]{huang2023robust}
\bibinfo{author}{\bibfnamefont{Z.}~\bibnamefont{Huang}},
  \bibinfo{author}{\bibfnamefont{E.}~\bibnamefont{Gull}}, \bibnamefont{and}
  \bibinfo{author}{\bibfnamefont{L.}~\bibnamefont{Lin}},
  \bibinfo{journal}{Physical Review B} \textbf{\bibinfo{volume}{107}},
  \bibinfo{pages}{075151} (\bibinfo{year}{2023}{\natexlab{b}}).

\bibitem[{\citenamefont{Bao et~al.}(2002)\citenamefont{Bao, Jin, and
  Markowich}}]{bao2002time}
\bibinfo{author}{\bibfnamefont{W.}~\bibnamefont{Bao}},
  \bibinfo{author}{\bibfnamefont{S.}~\bibnamefont{Jin}}, \bibnamefont{and}
  \bibinfo{author}{\bibfnamefont{P.~A.} \bibnamefont{Markowich}},
  \bibinfo{journal}{Journal of Computational Physics}
  \textbf{\bibinfo{volume}{175}}, \bibinfo{pages}{487} (\bibinfo{year}{2002}).

\bibitem[{\citenamefont{Kloeden et~al.}(1992)\citenamefont{Kloeden, Platen,
  Kloeden, and Platen}}]{kloeden1992stochastic}
\bibinfo{author}{\bibfnamefont{P.~E.} \bibnamefont{Kloeden}},
  \bibinfo{author}{\bibfnamefont{E.}~\bibnamefont{Platen}},
  \bibinfo{author}{\bibfnamefont{P.~E.} \bibnamefont{Kloeden}},
  \bibnamefont{and} \bibinfo{author}{\bibfnamefont{E.}~\bibnamefont{Platen}},
  \emph{\bibinfo{title}{Stochastic differential equations}}
  (\bibinfo{publisher}{Springer}, \bibinfo{year}{1992}).

\bibitem[{\citenamefont{Biele and D’Agosta}(2012)}]{biele2012stochastic}
\bibinfo{author}{\bibfnamefont{R.}~\bibnamefont{Biele}} \bibnamefont{and}
  \bibinfo{author}{\bibfnamefont{R.}~\bibnamefont{D’Agosta}},
  \bibinfo{journal}{Journal of Physics: Condensed Matter}
  \textbf{\bibinfo{volume}{24}}, \bibinfo{pages}{273201}
  (\bibinfo{year}{2012}).

\bibitem[{\citenamefont{Kaye and UR~Strand}(2023)}]{kaye2023fast}
\bibinfo{author}{\bibfnamefont{J.}~\bibnamefont{Kaye}} \bibnamefont{and}
  \bibinfo{author}{\bibfnamefont{H.}~\bibnamefont{UR~Strand}},
  \bibinfo{journal}{Advances in Computational Mathematics}
  \textbf{\bibinfo{volume}{49}}, \bibinfo{pages}{63} (\bibinfo{year}{2023}).

\bibitem[{\citenamefont{Eckstein and
  Werner}(2010)}]{eckstein2010nonequilibrium}
\bibinfo{author}{\bibfnamefont{M.}~\bibnamefont{Eckstein}} \bibnamefont{and}
  \bibinfo{author}{\bibfnamefont{P.}~\bibnamefont{Werner}},
  \bibinfo{journal}{Physical Review B} \textbf{\bibinfo{volume}{82}},
  \bibinfo{pages}{115115} (\bibinfo{year}{2010}).

\bibitem[{\citenamefont{Haertle and Millis}(2014)}]{haertle2014formation}
\bibinfo{author}{\bibfnamefont{R.}~\bibnamefont{Haertle}} \bibnamefont{and}
  \bibinfo{author}{\bibfnamefont{A.}~\bibnamefont{Millis}},
  \bibinfo{journal}{Physical Review B} \textbf{\bibinfo{volume}{90}},
  \bibinfo{pages}{245426} (\bibinfo{year}{2014}).

\end{thebibliography}

\onecolumngrid
\appendix

\section{Derivation of nMQME from nMSSE}
\label{appendix:QMEderivation}
We fill in the details of deriving the nMQME \cref{eq:QME}.
Recall that from \cref{eq:psi_expansion}
$$
\Psi^{(2)}_I(x,t) =-\mathrm{i} \left(\int_{0}^t\hat H_{\mathrm{int},I}^{(2)}(t_1)\mathrm{d}t_1\right)\Psi_I(x,0)  - \int_{0}^t\mathrm{d}t_1 \int_{0}^{t_1}\mathrm{d}t_2\hat H^{(1)}_{\mathrm{int},I}(t_1)H^{(1)}_{\mathrm{int},I}(t_2)\Psi_I(x,0).$$
Therefore
\begin{equation}
 \mathbb E \left(  \Psi^{(2)}_I(x,t) \right)=-\mathrm{i} \left(\int_{0}^t\hat H_{\mathrm{int},I}^{(2)}(t_1)\mathrm{d}t_1\right)\Psi_I(x,0)  - \int_{0}^t\mathrm{d}t_1 \int_{0}^{t_1}\mathrm{d}t_2\mathbb E\left(\hat H^{(1)}_{\mathrm{int},I}(t_1)H^{(1)}_{\mathrm{int},I}(t_2)\right)\Psi_I(x,0).
\end{equation}
Since $\mathbb{E}\left(\xi_i(t) \xi_{i^{\prime}}\left(t^{\prime}\right)\right)=0$, 
then
$$
\begin{aligned}
\mathbb E\left(\hat H^{(1)}_{\mathrm{int},I}(t_1)H^{(1)}_{\mathrm{int},I}(t_2)\right) = -\frac{|V(x)|^2}{4} &\left(\mathbb E(\xi_2(t_1)\xi_2(t_2))\mathbf I + \mathbb E(\xi_1(t_1)\xi_1(t_2))\mathbf I \right.
\\
&\left.+ \mathbb E(\xi_2(t_1)\xi_1(t_2))\sigma_x\sigma_y + \mathbb E(\xi_1(t_1)\xi_2(t_2))\sigma_y\sigma_x
\right) =0.
\end{aligned}
$$
Therefore
$$
\mathbb E \left(  \Psi^{(2)}_I(x,t) \right)=-\mathrm{i} \left(\int_{0}^t\hat H_{\mathrm{int},I}^{(2)}(t_1)\mathrm{d}t_1\right)\Psi_I(x,0).
$$
Taking derivatives with respect to $t$,
we have
$$
\begin{aligned}
    \frac{\mathrm{d}}{\mathrm{d}t} \mathbb E \left(  \Psi^{(2)}_I(x,t) \right) &= -\mathrm i \hat H_{\mathrm{int},I}^{(2)}(t)\Psi_I(x,0)\\& =
 \int_{0}^{t} \mathrm d \tau
 \left(
 \begin{array}{cc}
 c^-(\tau)V(x)\mathrm e^{-\mathrm i\hat h_1\tau}V(x)      &  \\
      & c^+(\tau)V(x)\mathrm e^{-\mathrm i\hat h_0\tau}V(x) 
 \end{array}
 \right)\left(\begin{array}{c}
     \psi_0(x,t-\tau)  \\
     \psi_1(x,t-\tau)
 \end{array}\right)\\& = 
 \int_{0}^{t} \mathrm d \tau
 \left(
 \begin{array}{cc}
 c^-(\tau)V(x)\mathrm e^{-\mathrm i\hat h_1\tau}V(x)      &  \\
      & c^+(\tau)V(x)\mathrm e^{-\mathrm i\hat h_0\tau}V(x) 
 \end{array}
 \right)\left(\begin{array}{c}
     \psi_0(x,0)  \\
     \psi_1(x,0)
 \end{array}\right)+ O(\lambda).
\end{aligned}$$
Another ingredient we need is $\mathbb E\left(|\Psi_I^{(1)}\rangle\langle \Psi_I^{(1)}|\right)$. Again recall from \cref{eq:psi_expansion} that 
$$
\Psi_I^{(1)}(x, t)=-\frac{ V(x)}{2}\left(\int_0^t 
\left(\xi_2(t_1) \sigma_x+\xi_1(t_1) \sigma_y\right)
\mathrm{d} t_1\right) \Psi_I(x, 0).
$$
Then
$$
\begin{aligned}
    \mathbb E\left(\langle x|\Psi_I^{(1)}\rangle\langle \Psi_I^{(1)}|x^{\prime}\rangle\right) = \frac{V(x)V(x^{\prime})}{4}\int_0^t\int_0^t \sigma_x \hat{\rho}_I^{(0)}(x,x^{\prime})\sigma_x \mathbb{E}\left(\xi_2(t_1)\xi_2^{*}(t_2)\right)
+ \sigma_y \hat{\rho}_I^{(0)}(x,x^{\prime})\sigma_y\mathbb{E}\left(\xi_1(t_1)\xi_1^{*}(t_2)\right) \\ +
\sigma_y \hat{\rho}_I^{(0)}(x,x^{\prime})\sigma_x \mathbb{E}\left(\xi_1(t_1)\xi_2^{*}(t_2)\right)
+ \sigma_x \hat{\rho}_I^{(0)}(x,x^{\prime})\sigma_y \mathbb{E}\left(\xi_2(t_1)\xi_1^{*}(t_2)\right)\mathrm{d}t_1\mathrm{d}t_2
\end{aligned}$$
Further simplifications give us

$$
\begin{aligned}
    \mathbb E\left(\langle x|\Psi_I^{(1)}\rangle\langle \Psi_I^{(1)}|x^{\prime}\rangle\right) = V(x)V(x^{\prime})\int_0^t\int_0^t 
    \left(
    \begin{array}{cc}
        \rho^{(0)}_{I,11}(x,x^{\prime})c^+(t_2-t_1) &  \\
         & \rho^{(0)}_{I,00}(x,x^{\prime})c^-(t_2-t_1)
    \end{array}
    \right)
    \mathrm{d}t_1\mathrm{d}t_2
\end{aligned}$$

Now we are ready to derive the master equation.
Since $\rho_I(t)=\mathbb{E} \hat{\rho}_I(t)$, $\mathbb{E} \hat{\rho}_I^{(0)}$ is a fixed value, 
$\mathbb{E} \hat{\rho}_I^{(1)}=0$, therefore $\frac{\mathrm{d}\rho_I(t)}{\mathrm{d}t}=\frac{\lambda^2}{\varepsilon^2}\frac{\mathrm{d}}{\mathrm{d}t} \mathbb{E} \hat{\rho}_I^{(2)}(t)+O(\lambda^3)$. We have
$$
\begin{aligned}
    \frac{\mathrm{d} \rho_I(x,x^{\prime},t)}{\mathrm{d} t} &= \frac{\lambda^2}{\varepsilon^2} \left(\langle x|\Psi_I^{(0)}\rangle \frac{\mathrm{d}}{\mathrm{d} t}\mathbb{E}\left(\langle \Psi_I^{(2)}|x^{\prime}\rangle\right) + \frac{\mathrm{d}}{\mathrm{d} t}\mathbb{E}\left(\langle x|\Psi_I^{(2)}\rangle\right)\langle \Psi_I^{(0)}|x^{\prime}\rangle+\frac{\mathrm{d}}{\mathrm{d} t}\mathbb E\left(\langle x|\Psi_I^{(1)}\rangle\langle \Psi_I^{(1)}|x^{\prime}\rangle\right)\right) + O(\lambda^3) \\
    & = \frac{\lambda^2}{\varepsilon^2} 
    \left(
         \int_0^t\mathrm{d}\tau\left(
 \begin{array}{cc}
  c^-(\tau)V(x)\mathrm e^{-\mathrm i\hat h_1\tau}V(x)      &  \\
      & c^+(\tau)V(x)\mathrm e^{-\mathrm i\hat h_0\tau} V(x)
 \end{array}
 \right)\rho_I^{(0)}(x,x^{\prime})\right.\\
 &+ \left.
 \int_0^t V(x)V(x^{\prime})
    \left(
    \begin{array}{cc}
        \rho^{(0)}_{I,11}(x,x^{\prime})c^+(t-\tau) &  \\
         & \rho^{(0)}_{I,00}(x,x^{\prime})c^-(t-\tau)
    \end{array}
    \right)
    \mathrm{d}\tau
 \right)+ h.c. + O(\lambda^3)
 \\
    & = \frac{\lambda^2}{\varepsilon^2} 
    \left(
        \int_0^t \mathrm{d}\tau\left(
 \begin{array}{cc}
  c^-(\tau)V(x)\mathrm e^{-\mathrm i\hat h_1\tau}V(x)      &  \\
      & c^+(\tau)V(x)\mathrm e^{-\mathrm i\hat h_0\tau} V(x)
 \end{array}
 \right)\rho_I(x,x^{\prime},t-\tau)\right.\\
 &+\left. 
 \int_0^t V(x)V(x^{\prime})
    \left(
    \begin{array}{cc}
        \rho_{I,11}(x,x^{\prime},\tau)c^+(t-\tau) &  \\
         & \rho_{I,00}(x,x^{\prime},\tau)c^-(t-\tau)
    \end{array}
    \right)
    \mathrm{d}\tau
 \right) +h.c.+ O(\lambda^3).
\end{aligned}
$$

\end{document}